\magnification=1200
\hoffset=.0cm
\voffset=.0cm
\baselineskip=.55cm plus .55mm minus .55mm

%
%
%
\input amssym.def
\input amssym.tex
%
%
%
%


\font\grassettogreco=cmmib10
\font\scriptgrassettogreco=cmmib7
\font\scriptscriptgrassettogreco=cmmib10 at 5 truept
\textfont13=\grassettogreco
\scriptfont13=\scriptgrassettogreco
\scriptscriptfont13=\scriptscriptgrassettogreco


\font\sansserif=cmss10
\font\scriptsansserif=cmss10 at 7 truept
\font\scriptscriptsansserif=cmss10 at 5 truept
\textfont14=\sansserif
\scriptfont14=\scriptsansserif
\scriptscriptfont14=\scriptscriptsansserif
\def\sans{\fam=14}


\font\capital=rsfs10
\font\scriptcapital=rsfs10 at 7 truept
\font\scriptscriptcapital=rsfs10 at 5 truept
\textfont15=\capital
\scriptfont15=\scriptcapital
\scriptscriptfont15=\scriptscriptcapital
\def\scri{\fam=15}


\font\euler=eusm10
\font\scripteuler=eusm7
\font\scriptscripteuler=eusm5 
\textfont12=\euler
\scriptfont12=\scripteuler
\scriptscriptfont12=\scriptscripteuler
\def\eul{\fam=12}

%
%
%
%
%

%
%
%
%
%
%
\def\ref#1{\lbrack#1\rbrack}
%
%
%
%
%

\def\ind{{\rm ind}\hskip 1pt}

\def\deg{{\rm deg}\hskip 1pt}

\def\ker{{\rm ker}\hskip 1pt}

\def\SL{{\rm SL}\hskip 1pt}
\def\U{{\rm U}}

\def\Fun{{\rm Fun}\hskip 1pt}

\def\Gau{{\rm Gau}\hskip 1pt}

\def\Tor{{\rm Tor}\hskip 1pt}
\def\Class{{\rm Class}\hskip 1pt}
\def\Princ{{\rm Princ}\hskip 1pt}

\def\Conn{{\rm Conn}\hskip 1pt}
\def\Harm{{\rm Harm}\hskip 1pt}

\def\hst1{\hskip 1pt}

\def\ul#1{\underline{#1}{}}
%
%
%
%
%

\hrule\vskip.5cm
\hbox to 16.5 truecm{November 2003  \hfil DFUB 2003--7}
\hbox to 16.5 truecm{Version 3  \hfil hep-th/0311143}
\vskip.5cm\hrule
\vskip.9cm
\centerline{\bf FOUR DIMENSIONAL ABELIAN DUALITY}  
\centerline{\bf AND $\SL(2,\Bbb Z)$ ACTION IN}
\centerline{\bf THREE DIMENSIONAL CONFORMAL FIELD THEORY}
\vskip.4cm
\centerline{by}
\vskip.4cm
\centerline{\bf Roberto Zucchini}
\centerline{\it Dipartimento di Fisica, Universit\`a degli Studi di Bologna}
\centerline{\it V. Irnerio 46, I-40126 Bologna, Italy}
\centerline{\it and }
\centerline{\it INFN, sezione di Bologna}
\vskip.9cm
\hrule
\vskip.6cm
\centerline{\bf Abstract} 
Recently, Witten showed that there is a natural action of the group 
$\SL(2,\Bbb Z)$ on the space of 3 dimensional conformal field theories
with $\U(1)$ global symmetry and a chosen coupling of the symmetry current
to a background gauge field on a $3$--fold $N$. He further argued that, 
for a class of conformal field theories, in the nearly Gaussian limit, this 
$\SL(2,\Bbb Z)$ action may be viewed as a holographic image of the well--known 
$\SL(2,\Bbb Z)$ Abelian duality of a pure $\U(1)$ gauge theory on AdS--like 
$4$--folds $M$ bounded by $N$, as dictated by the AdS/CFT correspondence. 
However, he showed that explicitly only for the generator $T$; 
for the generator $S$, instead, his analysis remained conjectural. 
In this paper, we propose a solution of this problem. 
We derive a general holographic formula for the nearly Gaussian generating functional 
of the correlators of the symmetry current and, using this, we show that Witten's 
conjecture is indeed correct when $N=S^3$. We further identify a class of homology
$3$--spheres $N$ for which Witten's conjecture takes a particular simple form.

\vskip.4cm
\par\noindent


\vfill\eject

\vskip .3cm {\bf 1. Introduction and conclusions}\vskip .3cm

Following a recent paper by Witten, the interest in 3 dimensional 
conformal field theory has been revived. In \ref{1},
Witten considered the set ${\sans CFT}(N)$ of all conformal field theories on a given
spin Riemannian $3$--fold $N$, which have a $\U(1)$ global symmetry. A field theory
${\scri T}\in{\sans CFT}(N)$ is specified by four sets of data: $a)$ the field 
content; $b)$ the overall sign of the symmetry current $j$; $c)$ 
the free parameters entering in the multipoint correlators of $j$; $d)$
the gauge invariant coupling of $j$ to a background gauge field $a$. 
Thus, $\scri T$ is completely described by the generating functional of 
the correlators of $j$
$$
{\cal Z}_{\scri T}(a)
=\bigg\langle\exp\bigg(\sqrt{-1}\int_N a\wedge \star j\bigg)\bigg\rangle_{\scri T}.
\eqno(1.1)
$$
By the above assumptions, ${\cal Z}_{\scri T}(a)$ is gauge invariant, i. e. 
$$
{\cal Z}_{\scri T}(a+df)={\cal Z}_{\scri T}(a),
\eqno(1.2)
$$
where $f$ is any function.
Developing on original ideas of Kapustin and Strassler \ref{2}, Witten showed
that there is a natural action of the group $\SL(2,\Bbb Z)$ on ${\sans CFT}(N)$.
That is, three operations $S$, $T$ and $C$ can be defined on ${\sans CFT}(N)$ 
satisfying the algebra
$$
\eqalignno{
\vphantom{\int}
S^2&=C,&(1.3a)\cr
\vphantom{\int}
(ST)^3&=1,&(1.3b)\cr
\vphantom{\int}
C^2&=1.&(1.3c)\cr
}
$$ 
The action of $\SL(2,\Bbb Z)$ on ${\sans CFT}(N)$ is concretely defined 
by three operators $\widehat S$, $\widehat T$ and $\widehat C$
acting on the functionals ${\cal Z}_{\scri T}(a)$ according to 
$$
\eqalignno{
\vphantom{\int}
\widehat S {\cal Z}_{\scri T}(a)&={\cal Z}_{S{\scri T}}(a),&(1.4a)\cr
\vphantom{\int}
\widehat T {\cal Z}_{\scri T}(a)&={\cal Z}_{T{\scri T}}(a),&(1.4b)\cr
\vphantom{\int}
\widehat C {\cal Z}_{\scri T}(a)&={\cal Z}_{C{\scri T}}(a).&(1.4c)\cr
}
$$
Their explicit expressions, found by Witten in \ref{1} (see also \ref{3}),
can be cast as 
$$
\eqalignno{
\vphantom{\int}
\widehat S {\cal Z}_{\scri T}(a)
&=\int Db{\cal Z}_{\scri T}(b)\exp\bigg({\sqrt{-1}\over 2\pi}\int_Nb\wedge da\bigg),
&(1.5a)\cr
}
$$
$$
\eqalignno{
\vphantom{\int}
\widehat T {\cal Z}_{\scri T}(a)
&={\cal Z}_{\scri T}(a)\exp\bigg({\sqrt{-1}\over 4\pi}\int_Na\wedge da\bigg),&(1.5b)\cr
\vphantom{\int}
\widehat C {\cal Z}_{\scri T}(a)
&={\cal Z}_{\scri T}(-a).&(1.5c)\cr
}
$$
Note that $\widehat S$ involves the quantization of the field $a$. Its expression
is only formal. The normalization of the measure $Db$ is unspecified. Further,
gauge fixing is tacitly assumed. 

In \ref{1}, Witten considered the physically relevant case where $N=S^3$.
There is an interesting subset ${\sans CFT}_0(N)$ of 3 dimensional conformal 
field theories, in which the $\SL(2,\Bbb Z)$ action is concretely realized. 
These are the large $N_f$ limit of 3 dimensional field theories of $N_f$ fermions 
with $\U(1)$ symmetry, which were studied a long time ago \ref{4--7} as well as 
recently \ref{8,9} in connection with the low energy strong coupling regime of 
3 dimensional QED. 
For a ${\scri T}\in{\sans CFT}_0(N)$, the correlators of the current $j$ 
are nearly Gaussian. 
Conformal invariance and unitarity entail that ${\cal Z}_{\scri T}(a)$ is given 
by the exponential of a quadratic expression in $a$ depending on two real parameters 
$\tau_1\in\Bbb R$, $\tau_2\in\Bbb R_+$, which can be organized in a complex parameter 
$\tau=\tau_1+\sqrt{-1}\tau_2\in\Bbb H_+$. Since these parameters completely
characterize the field theory ${\scri T}$, we can parameterize 
${\scri T}\in{\sans CFT}_0(N)$ with $\tau\in\Bbb H_+$ 
and write ${\cal Z}_{\scri T}(a)$ as ${\cal Z}(a,\tau)$. Witten showed that 
the $\SL(2,\Bbb Z)$ action on ${\sans CFT}_0(N)$ reduces to the customary
$\SL(2,\Bbb Z)$ modular action on $\Bbb H_+$:
$$
\eqalignno{
\vphantom{\int}
S(\tau)&=-1/\tau,&(1.6a)\cr
\vphantom{\int}
T(\tau)&=\tau+1,&(1.6b)\cr
\vphantom{\int}
C(\tau)&=\tau.&(1.6c)\cr
}
$$
One thus has correspondingly
$$
\eqalignno{
\vphantom{\int}
\widehat S {\cal Z}(a,\tau)&={\cal Z}(a,-1/\tau),&(1.7a)\cr
\vphantom{\int}
\widehat T {\cal Z}(a,\tau)&={\cal Z}(a,\tau+1),&(1.7b)\cr
\vphantom{\int}
\widehat C {\cal Z}(a,\tau)&={\cal Z}(a,\tau).&(1.7c)\cr
}
$$ 
Witten's analysis has been generalized in \ref{10} to higher spin conserved currents.

The action $S(A,\tau)$ of a pure $\U(1)$ gauge theory on a compact spin $4$--fold $M$ 
without boundary is given by
$$
S(A,\tau)={\sqrt{-1}\tau_2\over 4\pi}\int_M F_A\wedge \star F_A
+{\tau_1\over 4\pi}\int_MF_A\wedge F_A,
\eqno(1.8)
$$
where $A$ is the gauge field and $\tau=\tau_1+\sqrt{-1}\tau_2\in\Bbb H_+$ with
$\tau_1=\theta/2\pi\in\Bbb R$, $\tau_2=2\pi/e^2\in\Bbb R_+$. 
So, the partition function 
${\cal Z}(\tau)$ is a certain function of $\tau\in\Bbb H_+$. $\SL(2,\Bbb Z)$ 
operators $\widehat S$, $\widehat T$ and $\widehat C$ on ${\cal Z}(\tau)$ 
can be defined by relations analogous to (1.7) 
(with $a$ suppressed). Under their action, ${\cal Z}(\tau)$ behaves as a modular form 
of weights $\chi+\eta\over 4$, $\chi-\eta\over 4$, where $\chi$ and $\eta$ are 
respectively the Euler characteristic and the signature invariant of $M$ 
\ref{11--15}, a property usually referred to as Abelian duality. If $M$ has 
a non empty boundary $\partial M$, then the partition function ${\cal Z}(a,\tau)$ 
depends also on the assigned tangent component $a$ of the gauge field $A$ at 
$\partial M$. One expects that operators $\widehat S$, $\widehat T$ 
and $\widehat C$ can still be defined as in (1.7), though Abelian duality, 
as defined above, does not hold in this more general situation. 

In \ref{1}, Witten argued that, when $N=S^3$, the $\SL(2,\Bbb Z)$ action on 
${\sans CFT}_0(N)$ may be viewed as a holographic image of the $\SL(2,\Bbb Z)$ 
Abelian duality of a pure $\U(1)$ gauge theory on $M=B^4$, as implied by the 
Euclidean AdS/CFT correspondence, inasmuch as the $\SL(2,\Bbb Z)$ operators 
of ${\sans CFT}_0(N)$ can be equated to their gauge theory counterparts.
(Here, $B^4$ is the conformally compactified 
hyperbolic $4$--ball and $S^3=\partial B^4$ carries the induced 
conformal structure). He also indicated that this result should hold true 
for a general $3$--fold $N$ bounding AdS--like $4$--folds $M$. 
However, he showed that explicitly only for the generator $T$; for the generator $S$, 
instead, his analysis remained to some extent conjectural. In this paper, we propose 
a solution of this problem, which sheds some light on the holographic correspondence 
between the boundary conformal field theory and the bulk gauge theory at least at the 
linearized level.

Having holography in mind, we study in some detail Abelian gauge theory on a general 
oriented compact $4$--fold $M$ with boundary. The geometric framework appropriate 
for its treatment is provided by the theory of {\it (relative) principal $\U(1)$ 
\it bundle, gauge transformations and connections}. 
While the notions of principal bundle, gauge transformation and connection
are familiar to many theoretical physicists, those of their relative counterparts 
are much less so. Roughly speaking, a relative principal bundle (respectively
gauge transformation, connection) is an ordinary bundle (respectively
gauge transformation, connection), satisfying a suitable form of vanishing 
boundary conditions on $\partial M$. The framework is mathematically elegant,
is {\it natural}, as it parallels closely that of the boundaryless case,
and leads quite straightforwardly to the identification of the appropriate form 
of {\it boundary conditions} of the relevant bulk fields. 

In this paper, assuming the validity of the holographic correspondence, we derive 
a general holographic formula for the nearly Gaussian generating functional 
${\cal Z}(a,\tau)$ of the correlators of the symmetry current on a given 
$3$--fold $N$. Next, for definiteness, we concentrate on the case where a single 
AdS--like $4$--fold $M$ bounded by $N$ is required by holography.
In this case, ${\cal Z}(a,\tau)$ is given simply by
$$
{\cal Z}(a,\tau)=\exp\left(\sqrt{-1}S(A_a,\tau)\right),
\eqno(1.9)
$$
where $A_a$ is the gauge field on $M$ satisfying Maxwell's equation 
whose tangent component at the boundary equals $a$.
We verify explicitly that (1.9) is indeed correct when $N=S^3$ and $M=B^4$.
We further obtain sufficient topological conditions, which should be obeyed  
by general $N$ and $M$ in order (1.9) to hold. 
Finally, we identify a class of homology $3$--spheres $N$ for which Witten's 
conjecture takes the simple form (1.9).

This paper is structured as follows. 
In sect. 2, we review the main properties of (relative) principal $\U(1)$ bundles,
(relative) gauge transformations and (relative) connection on manifolds with boundary. 
Many results that hold for manifolds without boundary generalize in a non trivial 
manner. 
In sect. 3, we recall the basic Green identities for general forms, discuss the 
various choices of boundary conditions and describe the harmonic representation of 
(relative) cohomology.
In sect. 4, we compute the action $S(A,\tau)$ of eq. (1.8) for a $4$--fold 
$M$ with boundary as a functional of the tangential boundary component of 
the gauge field.
In sect. 5, we confirm Witten's claim of \ref{1} that the nearly Gaussian 
generating functional ${\cal Z}(a,\tau)$ of the correlators of $j$ for  $N=S^3$ 
is indeed given by (1.9) with $M=B^4$. We further derive sufficient 
topological conditions under which (1.9) holds for a general $3$--fold $N$ bounding 
a $4$--fold $M$.
Finally, in sect. 6, we identify a class of homology $3$--spheres $N$ fulfilling 
those conditions.

\vskip .3cm {\bf 2. Principal bundles, gauge transformations,  
connections and cohomology}
\vskip .3cm 

Let $M$ be a manifold with boundary $\partial M$.
We denote by $\iota_\partial{}:\partial M\rightarrow M$ 
be the natural inclusion map. We further denote by 
$d$ the de Rham differential of $M$ and by $d_\partial$ that of $\partial M$. 

Below, we assume the reader has some familiarity with the basics of sheaf cohomology 
(see \ref{16,17} for background material). Relative cohomology 
is tacitly assumed to mean relative cohomology of $M$ modulo $\partial M$.
For a sheaf of Abelian groups $\scri S$ over $M$, we denote by 
$H^p(M,{\scri S})$ the absolute $p$--th sheaf cohomology group
of $\scri S$ and by $H^p(M,\partial M,{\scri S})$ 
the relative $p$--th sheaf cohomology group of $\scri S$.
For an Abelian group $G$, $G$ denotes the associated constant sheaf 
on $M$. For an Abelian Lie group $G$, $\underline{G}$ denotes 
the sheaf of germs of smooth $G$ valued functions on $M$.
In this paper, $G$ will be one of the groups $\Bbb Z$, $\Bbb R$,
$\Bbb T$ (the circle group $\U(1)$).
We denote by $\Omega^p(M)$ the space of $p$--forms on $M$ and by 
$\Omega^p(M,\partial M)$ the space of relative $p$--forms on $M$.
We attach a subscript $\Bbb Z$ to denote the corresponding subsets of 
$p$--forms with integer (relative) periods. Analogous conventions hold for the 
absolute sheaf cohomology and the $p$--forms of $\partial M$.

Below, we occasionally use the \v Cech model of (relative) cohomology, which 
allows for a particularly simple and direct treatment. This requires the choice of a 
good open cover $\{O_i\}$ of $M$ such that $\{O_i\cap\partial M\}$ is a good open 
cover of $\partial M$. 

\vskip .3cm {\it  2.1. The basic exact sequence} 

The cohomological classification of the groups of principal $\underline{\Bbb T}$
bundles and the associated groups of gauge transformations is based on a standard
short exact sequence of sheaves:
$$
\matrix{&& &{}_i &&{}_e& &&\cr
0
&\!\!\!\longrightarrow\!\!\!& 
\Bbb Z
&\!\!\!\longrightarrow\!\!\!& 
\underline{\Bbb R}
&\!\!\!\longrightarrow\!\!\!& 
\underline{\Bbb T}
&\!\!\!\longrightarrow\!\!\!& 
0\cr
&&&&&&&&\cr
}
\eqno(2.1.1)
$$
where $i(n)=n$ for $n\in \Bbb Z$ and $e(x)=\exp(2\pi\sqrt{-1}x)$ for $x\in \Bbb R$.

As is well known, with (2.1.1) there is associated a long exact sequence of absolute
sheaf cohomology 
$$
\matrix{&&{}_{i_*}&&{}_{e_*}&&{}_\delta&&\cr
\longrightarrow\!\!\!\!& 
H^p(M,\Bbb Z)
&\!\!\!\!\longrightarrow\!\!\!\!& 
H^p(M,\underline{\Bbb R})
&\!\!\!\!\longrightarrow\!\!\!\!& 
H^p(M,\underline{\Bbb T})
&\!\!\!\!\longrightarrow\!\!\!\!& 
H^{p+1}(M,\Bbb Z)
&\!\!\!\!\longrightarrow.\cr
&&&&&&&&\cr
}
\eqno(2.1.2)
$$
The sheaf $\underline{\Bbb R}$ is fine and thus acyclic. 
Therefore, above, $H^p(M,\underline{\Bbb R})=0$ for $p>0$.
$H^0(M,\underline{\Bbb R})=\Fun(M,\Bbb R)$, the real valued functions on $M$.

Similarly, with (2.1.1) there is also associated a long exact sequence of 
relative sheaf cohomology 
$$
\matrix{&&{}_{i_*}&&{}_{e_*}&&{}_\delta&&\cr
\longrightarrow\!\!\!\!& 
H^p(M,\partial M,\Bbb Z)
&\!\!\!\!\longrightarrow\!\!\!\!& 
H^p(M,\partial M,\underline{\Bbb R})
&\!\!\!\!\longrightarrow\!\!\!\!& 
H^p(M,\partial M,\underline{\Bbb T})
&\!\!\!\!\longrightarrow\!\!\!\!& 
H^{p+1}(M,\partial M,\Bbb Z)
&\!\!\!\!\longrightarrow\!.\cr
&&&&&&&&\cr
}
\eqno(2.1.3)
$$
The fine sheaf $\underline{\Bbb R}$ is acyclic also in relative cohomology. 
Therefore, above, $H^p(M,\partial M,\underline{\Bbb R})=0$ for $p>0$.
$H^0(M,\partial M,\underline{\Bbb R})=\Fun(M,\partial M,\Bbb R)$,
the real valued functions on $M$ which vanish on $\partial M$.

For $\partial M$, one has a long exact sequence of cohomology analogous to 
(2.1.2):
$$
\matrix{&&{}_{i_{\partial *}}&&{}_{e_{\partial *}}&&{}_{\delta_\partial}&&\cr
\longrightarrow\!\!\!\!& 
H^p(\partial M,\Bbb Z)
&\!\!\!\!\longrightarrow\!\!\!\!& 
H^p(\partial M,\underline{\Bbb R})
&\!\!\!\!\longrightarrow\!\!\!\!& 
H^p(\partial M,\underline{\Bbb T})
&\!\!\!\!\longrightarrow\!\!\!\!& 
H^{p+1}(\partial M,\Bbb Z)
&\!\!\!\!\longrightarrow.\cr
&&&&&&&&\cr
}
\eqno(2.1.4)
$$
where $H^p(\partial M,\underline{\Bbb R})=0$ for $p>0$ and 
$H^0(\partial M,\underline{\Bbb R})=\Fun(\partial M,\Bbb R)$.

In addition to the above long exact sequences, we have the absolute/relative
cohomology long exact sequence
$$
\matrix{
&&{}_{}&&{}_{\iota_\partial{}_*}&&{}_\delta&&\cr  
\longrightarrow\!\!\!\!& 
H^p(M,\partial M,{\scri S}) 
&\!\!\!\!\longrightarrow\!\!\!\!& 
H^p(M,{\scri S})
&\!\!\!\!\longrightarrow\!\!\!\!& 
H^p(\partial M,{\scri S})
&\!\!\!\!\longrightarrow\!\!\!\!& 
H^{p+1}(M,\partial M,{\scri S})
&\!\!\!\!\longrightarrow,\cr
&&&&&&&&\cr}
\eqno(2.1.5)
$$
where $\scri S$ is one of the sheaves $\Bbb Z$, $\underline{\Bbb R}$,
$\underline{\Bbb T}$. This sequence is compatible with the sequences 
(2.1.2)--(2.1.4): the sequences (2.1.2)--(2.1.4) and (2.1.5)
can be arranged in a two dimensional commutative diagram with exact 
rows and columns.

\vskip .3cm {\it  2.2. Principal $\Bbb T$ bundles} 

Let $\Princ(M)$ be the group of principal $\Bbb T$ bundles over $M$.
(Below, we do not distinguish between isomorphic bundles).

There is a natural isomorphism
$$
\Princ(M)\cong H^1(M,\underline{\Bbb T}),
\eqno(2.2.1)
$$  
defined as follows. Let $P\in\Princ(M)$. Consider a set of local trivializations 
of $P$. The matching of the trivializations of $P$ is described by a 
$\underline{\Bbb T}$ \v Cech $1$--cochain $(T_{ij})$ on $M$. As is well known, 
the topology of $P$ requires that $(T_{ij})$ is a $1$--cocycle:
$T_{jk}T_{ik}{}^{-1}T_{ij}=1$. If we change the set of local trivializations used, 
the cocycle $(T_{ij})$ gets replaced by a cohomologous cocycle 
$(T'{}_{ij})$: $T'{}_{ij}=T_{ij}V_jV_i{}^{-1}$, for some 
$\underline{\Bbb T}$ \v Cech $0$--cochain $(V_i)$. Thus, with 
$P$, there is associated the cohomology class $[(T_{ij})]\in
H^1(M,\underline{\Bbb T})$. The resulting map $\Princ(M)\rightarrow 
H^1(M,\underline{\Bbb T})$ is easily shown to be an isomorphism.

From the long exact cohomology sequence (2.1.2), we obtain an isomorphism 
$H^1(M,\underline{\Bbb T})$ $\cong H^2(M,\Bbb Z)$. So, taking (2.2.1) into account, 
we have the isomorphism
$$
\matrix{
&{}_c&\cr
\Princ(M)\!\!\!\!&\cong & \!\!\!\! H^2(M,\Bbb Z).\cr
&&\cr
}
\eqno(2.2.2)
$$
$c$ assigns to a bundle $P$ its Chern class $c(P)$. The preimage by $c$ of the torsion
subgroup $\Tor H^2(M,\Bbb Z)$ of $H^2(M,\Bbb Z)$ is the subgroup $\Princ_0(M)$ 
of $\Princ(M)$ of flat principal $\Bbb T$ bundles.

The notion of relative principal bundle is a refinement of that of principal bundle.
A relative principal $\Bbb T$ bundle $(P,t)$ on $M$ consists of a principal 
$\Bbb T$ bundle $P$ on $M$ such that the principal bundle $\iota_\partial{}^*P$ 
on $\partial M$ is trivial and a trivialization 
$t:\iota_\partial{}^*P\rightarrow \partial M\times\Bbb T$. 
The relative principal $\Bbb T$ bundles form a group $\Princ(M,\partial M)$. 

The cohomological description of the group of relative principal bundles parallels
that of the group of principal bundles, but cohomology is replaced everywhere
by relative cohomology, as we show next. 

There is a natural isomorphism
$$
\Princ(M,\partial M)\cong H^1(M,\partial M,\underline{\Bbb T}),
\eqno(2.2.3)
$$
analogous to (2.2.1), defined as follows. Let $(P,t)\in\Princ(M,\partial M)$. 
Consider a set of local trivializations of $P$ and the induced 
set of local trivializations of $\iota_\partial{}^*P$. 
The matching of the trivializations of $P$ is described by a $\Bbb T$ \v Cech 
$1$--cochain $(T_{ij})$ on $M$, as before, while the matching of the induced 
trivializations of $\iota_\partial{}^*P$ and the trivialization 
$t$ is described by a $\Bbb T$ \v Cech 
$0$--cochain $(t_i)$ on $\partial M$. These \v Cech data form a relative 
$\underline{\Bbb T}$ \v Cech $1$--cochain $(T_{ij},t_i)$ on $M$. 
The topology of $P$ and $t$ requires that $(T_{ij},t_i)$ is a relative 
$1$--cocycle: $T_{jk}T_{ik}{}^{-1}T_{ij}=1$, as before, and further 
$\iota_\partial{}^*T_{ij}=t_jt_i{}^{-1}$.
If we change the set of local trivializations used, the relative cocycle 
$(T_{ij},t_i)$ gets replaced by a cohomologous relative cocycle $(T'{}_{ij},t'{}_i)$: 
$T'{}_{ij}=T_{ij}V_jV_i{}^{-1}$, as before, and further
$t'{}_i=t_i\iota_\partial{}^*V_i$, for some relative $\underline{\Bbb T}$ \v Cech 
$0$--cochain $(V_i)$ on $M$. Thus, with $(P,t)$, there is associated the relative 
cohomology class $[(T_{ij},t_i)]\in H^1(M,\partial M,\underline{\Bbb T})$. 
The resulting map $\Princ(M,\partial M)\rightarrow 
H^1(M,\partial M,\underline{\Bbb T})$ is easily shown to be an isomorphism.

From the long exact relative cohomology sequence (2.1.3), we obtain an isomorphism 
$H^1(M,\partial M,\underline{\Bbb T})\cong H^2(M,\partial M,\Bbb Z)$.
So, taking (2.2.3) into account, we have the isomorphism
$$
\matrix{
&{}_c&\cr
\Princ(M,\partial M)\!\!\!\!&\cong & \!\!\!\! H^2(M,\partial M,\Bbb Z).\cr
&&\cr
}
\eqno(2.2.4)
$$
$c$ assigns to a relative bundle $(P,t)$ its relative Chern class $c(P,t)$.
The preimage by $c$ of the torsion subgroup $\Tor H^2(M,\partial M,\Bbb Z)$ of 
$H^2(M,\partial M,\Bbb Z)$ is the subgroup $\Princ_0(M,\partial M)$ 
of $\Princ(M,\partial M)$ of flat relative principal $\Bbb T$ bundles. 

Of course, we can describe the group $\Princ(\partial M)$ of principal $\Bbb T$ 
bundles on $\partial M$ in precisely the same way as we did for the group 
$\Princ(M)$ of principal $\Bbb T$ bundles on $M$. So, the isomorphisms
(2.2.1), (2.2.2) hold with $M$, $c$ replaced by $\partial M$, $c_\partial$.
The preimage by $c_\partial$ of $\Tor H^2(\partial M,\Bbb Z)$ is the subgroup 
$\Princ_0(\partial M)$ of flat principal $\Bbb T$ bundles.

\vskip .3cm {\it  2.3. Gauge transformation group} 

Let $P\in\Princ(M)$ be a principal $\Bbb T$ bundle. 
We denote by $\Gau(P)$ the group of gauge transformations of $P$.
  
There is a natural isomorphism
$$
\Gau(P)\cong H^0(M,\underline{\Bbb T}),
\eqno(2.3.1)
$$
defined as follows. Let $U\in\Gau(P)$. Consider a set of local trivializations 
of $P$. The local representatives $U_i$ of $U$ in the various trivializing domains
form a $\underline{\Bbb T}$ \v Cech $0$--cochain $(U_i)$ on $M$. As is well known, 
the global definedness of $U$ requires that $(U_i)$ is a $0$--cocycle:
$U_jU_i{}^{-1}=1$. If we change the set of local trivializations used, the cocycle 
$(U_i)$ is left unchanged. Thus, with $U$ there is associated a  cohomology class 
$[(U_i)]\in H^0(M,\underline{\Bbb T})$. The resulting map $\Gau(P)\rightarrow 
H^0(M,\underline{\Bbb T})$ is easily shown to be an isomorphism.
As $\Gau(P)$ does not depend on $P$, we shall also denote it as 
$\Gau(M)$. 

From the long exact cohomology sequence (2.1.2), we have a homomorphism
$H^0(M,\underline{\Bbb T})$ $\rightarrow H^1(M,\Bbb Z)$ with kernel 
$H^0(M,\underline{\Bbb R})/i_*H^0(M,{\Bbb Z})$. Taking (2.3.1)
into account, we obtain a homomorphism 
$\Gau(M)\rightarrow H^1(M,\Bbb Z)$ with kernel 
$$
\Gau_c(M)\cong H^0(M,\underline{\Bbb R})/i_*H^0(M,{\Bbb Z}).
\eqno(2.3.2)
$$
Hence, we have a short exact sequence
$$
\matrix{&& && &{}_{q}& &&\cr
0&\!\!\!\longrightarrow\!\!\!& \Gau_c(M)
&\!\!\!\longrightarrow\!\!\!& \Gau(M)
&\!\!\!\longrightarrow\!\!\!& H^1(M,\Bbb Z)  &\!\!\!\longrightarrow\!\!\!& 0.\cr
&&&&&&&&\cr
}
\eqno(2.3.3)
$$
$q$ assigns to a gauge transformation $U$ its characteristic class $q(U)$.
The preimage by $q$ of the torsion subgroup $\Tor H^1(M,\Bbb Z)$ 
of $H^1(M,\Bbb Z)$ is the subgroup $\Gau_0(M)$ of $\Gau(M)$ of flat 
gauge transformations.

A relative gauge transformation $U$ of the relative principal $\Bbb T$ bundle 
$(P,t)\in\Princ(M,$ $\partial M)$ is a gauge transformation $U$ of 
the underlying bundle $P$ whose pull--back $\iota_\partial{}^*U$
equals the pull--back by $t$ of the trivial gauge transformations $1$ of the bundle
$\partial M\times{\Bbb T}$, $\iota_\partial{}^*U=t^*1$. 
(Recall that $\iota_\partial{}^*U$ is a gauge transformation of the bundle 
$\iota_\partial{}^*P$ and that 
$t:\iota_\partial{}^*P\rightarrow\partial M\times\Bbb T$
is a trivialization).
We denote by $\Gau(P,t)$ the group of relative gauge transformations of $(P,t)$.

The cohomological description of the group of relative gauge transformations parallels
that of the group of gauge transformations, but cohomology is replaced everywhere
by relative cohomology, as happens for the relative principal bundles.

There is a natural isomorphism
$$
\Gau(P,t)\cong H^0(M,\partial M,\underline{\Bbb T}),
\eqno(2.3.4)
$$
defined as follows. Let $U\in\Gau(P,t)$. Consider a set of local trivializations 
of $P$ and the induced set of local trivializations of $\iota_\partial{}^*P$. 
The local representatives $U_i$ of $U$ in the various trivializing domains
form a relative $\underline{\Bbb T}$ \v Cech $0$--cochain $(U_i)$ on $M$. 
The global definedness of $U$ and the triviality of $\iota_\partial{}^*U$ 
require that $(U_i)$ is actually a relative $0$--cocycle:
$U_jU_i{}^{-1}=1$, $\iota_\partial{}^*U_i=1$. 
If we change the set of local trivializations used, the relative cocycle 
$(U_i)$ is left unchanged. Thus, with $U$ there is associated a relative 
cohomology class $[(U_i)]\in H^0(M,\partial M,\underline{\Bbb T})$. 
The resulting map $\Gau(P,t)\rightarrow H^0(M,\partial M,\underline{\Bbb T})$ 
is easily shown to be an isomorphism.
As $\Gau(P,t)$ does not depend on $(P,t)$, we shall use also denote it as 
$\Gau(M,\partial M)$. 

From the long exact relative cohomology sequence (2.1.3), we have a homomorphism
$H^0(M,\partial M,\underline{\Bbb T})\rightarrow H^1(M,\partial M,\Bbb Z)$ 
with kernel $H^0(M,\partial M,\underline{\Bbb R})/i_*H^0(M,\partial M,{\Bbb Z})$.
On account of (2.3.4), this yields a homomorphism
$\Gau(M,\partial M)\rightarrow H^1(M,\partial M,\Bbb Z)$ with kernel 
$$
\Gau_c(M,\partial M)\cong 
H^0(M,\partial M,\underline{\Bbb R})/i_*H^0(M,\partial M,{\Bbb Z}).
\eqno(2.3.5)
$$
Hence, we have a short exact sequence
$$
\matrix{&& && &{}_{q}& &&\cr
0&\!\!\!\longrightarrow\!\!\!& \Gau_c(M,\partial M)
&\!\!\!\longrightarrow\!\!\!& \Gau(M,\partial M)
&\!\!\!\longrightarrow\!\!\!& H^1(M,\partial M,\Bbb Z)  
&\!\!\!\longrightarrow\!\!\!& 0.\cr
&&&&&&&&\cr
}
\eqno(2.3.6)
$$
$q$ assigns to a relative gauge transformation $U$ its relative characteristic 
class $q(U)$. The preimage by $q$ of the torsion subgroup 
$\Tor H^1(M,\partial M,\Bbb Z)$ of $H^1(M,\partial M,\Bbb Z)$ is the subgroup 
$\Gau_0(M,\partial M)$ of $\Gau(M,\partial M)$ of flat relative gauge 
transformations.

Of course, we can describe the group $\Gau(\partial M)$ of gauge transformations
on $\partial M$ in precisely the same way as we did for the group 
$\Princ(M)$ of gauge transformations on $M$. So, the isomorphisms
(2.3.1), (2.3.2) and the short exact sequence (2.3.3)  
hold with $M$, $q$ replaced by $\partial M$, $q_\partial$.
The preimage by $q_\partial$ of $\Tor H^1(\partial M,\Bbb Z)$ is the 
subgroup $\Gau_0(\partial M)$ of flat gauge transformations.

\vskip .3cm {\it  2.4. Connections and gauge transformations} 

Let $P\in\Princ(M)$ be a principal $\Bbb T$ bundle. 
We denote by $\Conn(P)$ the affine space of connections of $P$. 

Let $A\in\Conn(P)$ be a connection of $P$. Suppose that $P$ is represented by the 
$\underline{\Bbb T}$ \v Cech $1$--cocycle $(T_{ij})$ with respect to a set of 
local trivializations. The local representatives $A_i$ 
of $A$ in the various trivializing 
domains constitute a $\Omega^1$ \v Cech $0$--cochain $(A_i)$ on $M$. 
As is well known, the global definedness of $A$ requires that the $0$--cochain 
$(A_i)$ satisfies the matching relation $A_j-A_i=-\sqrt{-1}T_{ij}{}^{-1}dT_{ij}$.
Let $(T_{ij})$, $(T'{}_{ij})$ be the cohomologous $\underline{\Bbb T}$
\v Cech $1$--cocycles corresponding to two different choices of local
trivializations of $P$, so that $T'{}_{ij}=T_{ij}V_jV_i{}^{-1}$, 
for some $\underline{\Bbb T}$ \v Cech $0$--cochain $(V_i)$. 
Then, the $0$--cochains $(A_i)$, $(A'{}_i)$ representing $A$ 
with respect to the two sets of local trivializations are related as
$A'{}_i=A_i-\sqrt{-1}V_i{}^{-1}dV_i$.

The curvature $F_A$ of a connection $A\in\Conn(P)$ is defined by 
$$
F_A=dA.
\eqno(2.4.1)
$$
This is a concise expression of the local relations $F_A|_{O_i}=dA_i$. 
The properties of $A$ listed in the previous paragraph ensure that  
$F_A$ does not depend on the chosen local trivializations. 
Therefore, $F_A\in\Omega^2(M)$ is a $2$--form.
$F_A$ is obviously closed:
$$
dF_A=0.
\eqno(2.4.2)
$$
Further, $F_A/2\pi$ has integer periods, that is 
$$
{1\over 2\pi}\int_SF_A\in \Bbb Z,
\eqno(2.4.3)
$$
for any singular $2$--cycle $S$. Recall that a singular $p$--cycle $X$ of $M$ is 
$p$--chain $X$ such that $bX=0$, $b$ being the singular boundary operator. 
Thus, $F_A/2\pi\in\Omega^2_{\Bbb Z}(M)$ and, so, it represents a class 
$x(P)$ of the integer lattice $H_{\Bbb Z}^2(M,\Bbb R)$ of $H^2(M,\Bbb R)$. Indeed, 
$x(P)$ is the image of the Chern class $c(P)$ of $P$
under the natural homomorphism $H^2(M,\Bbb Z)\rightarrow H^2(M,\Bbb R)$.
$x(P)$ vanishes precisely for the flat bundles $P\in\Princ_0(M)$.
The above statements can be shown in straightforward fashion using 
the \v Cech--de Rham cohomology double complex. 

A relative connection $A$ of a relative principal $\Bbb T$ bundle $(P,t)\in
\Princ(M,\partial M)$ is a connection $A$ of the underlying bundle $P$ 
whose pull--back $\iota_\partial{}^*A$ equals the pull--back by $t$
of the trivial connection $0$ of the bundle $\partial M\times {\Bbb T}$, 
$\iota_\partial{}^*A=t^*0$.
(Recall that $\iota_\partial{}^*A$ is a connection of the bundle 
$\iota_\partial{}^*P$ and that 
$t:\iota_\partial{}^*P\rightarrow\partial M\times\Bbb T$
is a trivialization).
We denote by $\Conn(P,t)$ the affine space of relative connections of $(P,t)$. 

Let $A\in\Conn(P,t)$ be a relative connection of $(P,t)$. 
Suppose that $(P,t)$ is represented by the relative $\underline{\Bbb T}$ 
\v Cech $1$--cocycle $(T_{ij},t_i)$ with respect to a set of local (induced) 
trivializations. 
The local representatives $A_i$ of $A$ in the various trivializing domains
form a relative $\Omega^1$ \v Cech $0$--cochain $(A_i)$. 
The global definedness of $A$ and the triviality of $\iota_\partial{}^*A$ 
require that the relative $0$--cochain $(A_i)$ satisfies the matching relation
$A_j-A_i=-\sqrt{-1}T_{ij}{}^{-1}dT_{ij}$, as before, and further that
$\iota_\partial{}^*A_i=-\sqrt{-1}t_i{}^{-1}d_\partial t_i$.
Let $(T_{ij},t_i)$, $(T'{}_{ij},t'{}_i)$ be the cohomologous relative 
$\underline{\Bbb T}$ \v Cech $1$--cocycles on $M$ corresponding to two different 
choices of local (induced) trivializations, so that
$T'{}_{ij}=T_{ij}V_jV_i{}^{-1}$, $t'{}_i=t_i\iota_\partial{}^*V_i$
for some relative $\underline{\Bbb T}$ \v Cech $0$--cochain $(V_i)$. 
Then, the relative $0$--cochains $(A_i)$, $(A'{}_i)$ representing $A$ 
with respect to the two sets of local (induced) trivializations are related 
again as $A'{}_i=A_i-\sqrt{-1}V_i{}^{-1}dV_i$.

The curvature $F_A$ of a relative connection $A\in\Conn(P,t)$ is simply the curvature 
$F_A$ of $A$ viewed as a connection of $P$ and is thus given by (2.4.1).
In addition, however, $F_A$ satisfies the boundary conditions
$$
\iota_\partial{}^*F_A=0.
\eqno(2.4.4)
$$
Thus, $F_A\in\Omega^2(M,\partial M)$ is a relative $2$--form.
$F_A$ still satisfies (2.4.2) and, so, is closed.
Further, $F_A/2\pi$ has integer relative periods, that is 
it satisfies (2.4.3) for any relative singular $2$--cycle $S$. 
Recall that a relative singular $p$--cycle $X$ is a 
singular $p$--chain $X$ such that $bX$ is supported in $\partial M$. 
Thus, $F_A/2\pi\in\Omega^2_{\Bbb Z}(M,\partial M)$ and, so, it represents 
a class $x(P,t)$ of the integer lattice $H_{\Bbb Z}^2(M,\partial M,\Bbb R)$ 
of $H^2(M,\partial M,\Bbb R)$. $x(P,t)$ is the image of the relative Chern class 
$c(P,t)$ of $(P,t)$ under the natural homomorphism $H^2(M,\partial M,\Bbb Z)
\rightarrow H^2(M,\partial M,\Bbb R)$. $x(P,t)$ vanishes precisely for the flat 
relative bundles $(P,t)\in\Princ_0(M,\partial M)$.
The above statements can be shown using the relative \v Cech--de Rham cohomology 
double complex. 

For a gauge transformation $U\in\Gau(M)$, we define 
$$
B_U=-\sqrt{-1}U^{-1}dU.
\eqno(2.4.5)
$$
As for the curvature of a connection, this is a concise expression of 
the local relations $B_U|_{O_i}=-\sqrt{-1}U_i^{-1}dU_i$, in which the right hand 
side does not depend on the chosen local trivialization. 
$B_U\in\Omega^1(M)$ is a $1$--form. $B_U$ is obviously closed:
$$
dB_U=0.
\eqno(2.4.6)
$$
Further, $B_U/2\pi$ has integer periods, that is 
$$
{1\over 2\pi}\int_S B_U\in \Bbb Z,
\eqno(2.4.7)
$$
for any singular $1$--cycle $S$. Thus, $B_U/2\pi\in\Omega^1_{\Bbb Z}(M)$
and, so, it represents a class $z(U)$ of the integer lattice
$H_{\Bbb Z}^1(M,\Bbb R)$ of $H^1(M,\Bbb R)$.
$z(U)$ is the image of the characteristic class $q(U)$ of $U$
under the natural homomorphism $H^1(M,\Bbb Z)\rightarrow H^1(M,\Bbb R)$
and vanishes precisely for the flat gauge transformations $U\in\Gau_0(M)$.

A gauge transformation $U\in\Gau(M)$ acts on a connection $A\in\Conn(P)$
of a principal $\Bbb T$ bundle $P$ as
$$
A^U=A+B_U.
\eqno(2.4.8)
$$
By (2.4.6), the curvature $F_A$ is gauge invariant:
$$
F_{A^U}=F_A.
\eqno(2.4.9)
$$
Note that the gauge transformation group is smaller than the invariance group of the 
curvature $F_A$. Indeed, $F_A$ is invariant under shifts of $A$ by an arbitrary closed
$1$--form.

With any relative gauge transformation $U\in\Gau(M,\partial M)$ there is associated 
a $1$--form $B_U$ defined as in (2.4.5). In addition, however, $B_U$ satisfies the 
boundary conditions
$$
\iota_\partial{}^*B_U=0.
\eqno(2.4.10)
$$
Thus, $B_U\in\Omega^1(M,\partial M)$ is a relative $1$--form.
$B_U$ still satisfies (2.4.6) and, so, is closed.
Further, $B_U/2\pi$ has integer relative periods, so that
it satisfies (2.4.7) for any relative singular $1$--cycle $S$. 
Thus, $B_U/2\pi\in\Omega^1_{\Bbb Z}(M,\partial M)$ and, so, 
it represents a class $z(U)$ of the integer lattice
$H_{\Bbb Z}^1(M,\partial M,\Bbb R)$ of $H^1(M,\partial M,\Bbb R)$.
$z(U)$ is the image of the relative characteristic class $q(U)$ of $U$
under the natural homomorphism $H^1(M,\partial M,\Bbb Z)\rightarrow 
H^1(M,\partial M,\Bbb R)$ and vanishes precisely for the flat relative 
gauge transformations $U\in\Gau_0(M,\partial M)$.

The action of a relative gauge transformation $U\in \Gau(M,\partial M)$
on a relative connection $A\in\Conn(P,t)$ of a relative principal 
$\Bbb T$ bundle $(P,t)$ is again given by (2.4.8). By (2.4.10), 
(2.4.4) is preserved.

We can describe the affine space $\Conn(P_\partial)$ of connections
of a bundle $P_\partial\in\Princ(\partial M)$ and the action of the 
gauge transformation group $\Gau(\partial M)$ on it in precisely the same 
way as we did for affine space $\Conn(P)$ of connections of a bundle 
$P\in\Princ(M)$ and the action of the gauge transformation group $\Gau(M)$ on it. 

\vskip .3cm {\it  2.5. Extendability of principal bundles and gauge transformations on 
$\partial M$ to $M$} 

Every bundle $P\in\Princ(M)$ yields by pull--back a bundle 
$P_\partial\in\Princ(\partial M)$, viz $P_\partial=\iota_\partial{}^*P$. 
The converse is however false: in general, not 
every bundle $P_\partial\in\Princ(\partial M)$ is the pull--back
of some bundle $P\in\Princ(M)$. When this does indeed happen, we say 
that $P_\partial$ is extendable to $M$. It is important to find out under which 
conditions a given bundle $P_\partial\in\Princ(\partial M)$ is extendable. 
To this end, consider the absolute/relative cohomology long exact sequence 
(2.1.5) with ${\scri S}={\Bbb Z}$. We can exploit the isomorphisms (2.2.4),  
(2.2.2) and its analogue for $\partial M$ to draw the commutative 
diagram
$$
\matrix{
&{}_\delta& &{}_{}& &{}_{\iota_\partial{}^*} & & {}_{\delta} & & \cr  
H^1(\partial M,\Bbb Z)&\!\!\!\longrightarrow\!\!\!  & H^2(M,\partial M,\Bbb Z) 
&\!\!\!\longrightarrow\!\!\!  & 
H^2(M,\Bbb Z)&\!\!\!\longrightarrow\!\!\!  & H^2(\partial M,\Bbb Z)&
\!\!\!\longrightarrow\!\!\!  & H^3(M,\partial M,\Bbb Z)\cr
&&{~\atop c}{~\atop\big\uparrow}~~&&{~\atop c}{~\atop\big\uparrow}~~&&
{~\atop c_\partial}{~\atop\big\uparrow}~~&&\cr
&& &{}_{}& &{}_{\iota_\partial{}^*} & & {}_{} & & \cr  
&& \Princ(M,\partial M) &\!\!\!\longrightarrow\!\!\!  & 
\Princ(M)&\!\!\!\longrightarrow\!\!\!  & \Princ(\partial M)&\cr
}
\eqno(2.5.1)
$$
in which the lines are exact and the vertical mappings are isomorphisms.
The interpretation of the second line is quite simple. The first mapping 
associates with every relative bundle $(P,t)\in\Princ(M,\partial M)$ 
the underlying bundle $P\in\Princ(M)$, the second associates with every bundle 
$P\in\Princ(M)$ its pull--back bundle $P_\partial=\iota_\partial{}^*P
\in\Princ(\partial M)$. This interpretation goes over to the first line. 

By the exactness of (2.5.1), a bundle $P_\partial\in\Princ(\partial M)$
is the pull--back of a bundle $P\in\Princ(M)$ if and only if 
$$
\delta(c_\partial(P_\partial))=0.
\eqno(2.5.2)
$$
Hence, the obstruction to the extendability of $P_\partial$ is a class
of $H^3(M,\partial M,\Bbb Z)$. When $P_\partial$ satisfies (2.5.2), $P_\partial$
is in general the pull--back of several bundles $P$ on $M$, i. e. $P_\partial$ 
has several extensions to $M$. Again, by the exactness of (2.5.1), its extensions are
parameterized by the group of relative bundles $\Princ(M,\partial M)$, 
in a generally non one--to--one fashion. The parameterization is one--to--one if 
$H^1(\partial M,\Bbb Z)=0$.

A similar analysis can be carried out for gauge transformations.
Every gauge transformation $U\in\Gau(M)$ yields by pull--back a gauge transformation 
$U_\partial\in\Gau(\partial M)$, viz $U_\partial=\iota_\partial{}^*P$. 
The converse is however false: in general, not 
every gauge transformation $U_\partial\in\Gau(\partial M)$ is the pull--back
of some gauge transformation $U\in\Gau(M)$. When this does indeed happen, we say 
that $U_\partial$ is extendable to $M$. As for principal bundles, 
it is important to find out under which conditions a given gauge 
transformation $U_\partial\in\Gau(\partial M)$ is extendable. 
In practice, this can be done only for 
gauge transformation classes. To this end, introduce the 
gauge transformation class groups 
$$
\Class(M)=\Gau(M)/\Gau_c(M),
\eqno(2.5.3)
$$
$$
\Class(M,\partial M)=\Gau(M,\partial M)/\Gau_c(M,\partial M)
\eqno(2.5.4)
$$
and $\Class(\partial M)$, which is given by (2.5.3) with $M$ replaced by $\partial M$.
Next, onsider again the absolute/relative cohomology long exact sequence 
(2.1.5) with ${\scri S}={\Bbb Z}$. We can exploit the short exact sequences
(2.3.3), (2.3.6) and its analogue for $\partial M$ to draw the commutative 
diagram
$$
\matrix{
&{}_\delta& &{}_{}& &{}_{\iota_\partial{}^*} & & {}_{\delta} & & \cr  
H^0(\partial M,\Bbb Z)&\!\!\!\longrightarrow\!\!\!  & H^1(M,\partial M,\Bbb Z) 
&\!\!\!\longrightarrow\!\!\!  & 
H^1(M,\Bbb Z)&\!\!\!\longrightarrow\!\!\!  & H^1(\partial M,\Bbb Z)&
\!\!\!\longrightarrow\!\!\!  & H^2(M,\partial M,\Bbb Z)\cr
&&{~\atop q}{~\atop\big\uparrow}~~&&{~\atop q}{~\atop\big\uparrow}~~&&
{~\atop q_\partial}{~\atop\big\uparrow}~~&&\cr
&& &{}_{}& &{}_{\iota_\partial{}^*} & & {}_{} & & \cr  
&& 
\Class(M,\partial M)
&\!\!\!\longrightarrow\!\!\!  & 
\Class(M)
&\!\!\!\longrightarrow\!\!\!& 
\Class(\partial M)&\cr
}
\eqno(2.5.5)
$$
in which the lines are exact and the vertical mappings are isomorphisms.
In the second line, 
the first mapping associates with every relative gauge transformation class
$[U]\in\Class(M,\partial M)$ the underlying gauge transformation class
$[U]\in\Class(M)$, the second associates with every gauge transformation 
class $[U]\in\Class(M)$ its pull--back gauge transformation class
$[U_\partial]=[\iota_\partial{}^*U]\in\Class(\partial M)$. 
This interpretation extends to the first line. 

By the exactness of (2.5.5), a gauge transformation class $[U_\partial]
\in\Class(\partial M)$ is the pull--back of a gauge transformation class
$[U]\in\Class(M)$ if and only if 
$$
\delta(q_\partial([U_\partial]))=0.
\eqno(2.5.6)
$$
Hence, the obstruction to the extendability of $[U_\partial]$ is a class
of $H^2(M,\partial M,\Bbb Z)$. When $[U_\partial]$ satisfies (2.5.6), $[U_\partial]$
is the pull--back of several gauge transformation classes $[U]\in\Class(M)$, i. e. 
$[U_\partial]$ has several extensions to $M$. Again, by the exactness of (2.5.5), 
its extensions are parameterized by the group of relative gauge transformations 
$\Gau(M,\partial M)$ in a non one--to--one fashion, since
$H^0(\partial M,\Bbb Z)\not=0$.

\vskip .3cm {\bf 3. Green identities, boundary conditions and harmonic forms}
\vskip .3cm 

Let $M$ be a compact oriented $m$--manifold with boundary $\partial M$.
Let $g$ be a metric on $M$. Let $g_\partial=\iota_\partial{}^*g$ be the induced 
metric on $\partial M$. We denote by $n$ the outward unit normal field to 
$\partial M$. Actually, $n$ is defined in a collar neighborhood of $\partial M$.

\vskip .3cm {\it  3.1. Basic Green identities} 

Let $d^*=(-1)^{mp+m+1}\star d\star$ denote the formal adjoint of $d$, 
with $p$ the form degree. The basic integral identity relating $d$ and $d^*$
is
$$
\int_Md\alpha\wedge\star\beta-\int_M\alpha\wedge\star d^*\beta
=\oint_{\partial M}\iota_\partial{}^*\alpha\wedge
\star_\partial\iota_\partial{}^*j(n)\beta,
\eqno(3.1.1)
$$
for $\alpha\in\Omega^p(M)$, $\beta\in\Omega^{p+1}(M)$, where $j(n)$ is the 
contraction operator associated with $n$.
\footnote{}{}\footnote{${}^1$}
{In local coordinates $j(n)\alpha_{i_1\cdots i_{p-1}}=n^i\alpha_{ii_1\cdots i_{p-1}}$ 
for any $p$--form $\alpha$.} 
(3.1.1) follows easily from Stokes' theorem upon using the identity
$\iota_\partial{}^*\star\beta=\star_\partial\iota_\partial{}^*j(n)\beta$.

From (3.1.1), a number of other basic integral identities follow.
Let $\Delta=dd^*+d^*d$ be the Hodge Laplacian. For $\alpha,~\beta\in\Omega^p(M)$, 
one has
$$
\eqalignno{
\int_M\alpha\wedge\star\Delta\beta-
\int_Md\alpha\wedge\star d\beta-\int_Md^*\alpha\wedge\star d^*\beta
&=\oint_{\partial M}\big[
\iota_\partial{}^*d^*\beta \wedge\star_\partial \iota_\partial{}^*j(n)\alpha
&(3.1.2)\cr
&\phantom{=\oint_{\partial M}}
-\iota_\partial{}^*\alpha \wedge\star_\partial \iota_\partial{}^*j(n)d\beta\big]
&\cr
}
$$
(1st Green identity) and
$$
\eqalignno{
\int_M\alpha\wedge\star\Delta\beta-\int_M\beta\wedge\star\Delta\alpha
&=\oint_{\partial M} \big[
\iota_\partial{}^*d^*\beta \wedge\star_\partial \iota_\partial{}^*j(n)\alpha
-\iota_\partial{}^*\alpha \wedge\star_\partial \iota_\partial{}^*j(n)d\beta~~~~~~~~ 
&(3.1.3)\cr
&\phantom{=\oint_{\partial M}}
-\iota_\partial{}^*d^*\alpha\wedge\star_\partial \iota_\partial{}^*j(n)\beta 
+\iota_\partial{}^*\beta  \wedge\star_\partial \iota_\partial{}^*j(n)d\alpha\big]
&\cr
}
$$
(2nd Green identity) \ref{18}. 

For any form degree $p$, a $p$--form Green operator $G$ of the Hodge Laplacian 
$\Delta$ is a symmetric distributional biform of bidegree $(p,p)$ on $M\times M$ 
satisfying the equation 
$$
\Delta G_x=\star\delta_x,
\eqno(3.1.4)
$$
for all $x\in M\setminus\partial M$, where $G_x$ is the $p$--form obtained by fixing 
one of the arguments of $G_x$ at $x$ and $\delta_x$ is the Dirac distribution 
for $p$--forms centered in $x$. The existence and uniqueness of Green operators
depend on the boundary conditions imposed.

A $p$--form $\alpha\in\Omega^p(M)$ is harmonic, if $\Delta\alpha=0$. 
The harmonic $p$--forms of $M$ form a subspace $\Harm^p(M)$
of $\Omega^p(M)$. 

Let $G$ be a Green operator of the Hodge Laplacian $\Delta$ on $p$--forms. 
Using the 2nd Green identity, one can show that, for any $\alpha\in\Harm^p(M)$, 
$$
\eqalignno{
\alpha(x)=(-1)^{p(m-p)}&\oint_{\partial M} \big[
\iota_\partial{}^*d^* G_x \wedge\star_\partial \iota_\partial{}^*j(n)\alpha
-\iota_\partial{}^*\alpha \wedge\star_\partial \iota_\partial{}^*j(n)d G_x
&(3.1.5)\cr
&\phantom{\oint_{\partial M}}
-\iota_\partial{}^*d^*\alpha\wedge\star_\partial \iota_\partial{}^*j(n) G_x
+\iota_\partial{}^* G_x\wedge\star_\partial \iota_\partial{}^*j(n)d\alpha
\big],&\cr
}
$$
for all $x\in M\setminus\partial M$. This expression simplifies to a considerable 
extent, when $\alpha$ and $G$ obey a suitable set of boundary conditions.
It is possible to extend the right hand side of (3.1.5) to $x\in\partial M$ by 
using a suitable limiting procedure.

\vskip .3cm {\it  3.2. Boundary conditions and Hilbert structures}

There are several relevant choices of boundary conditions. All these are combinations
of two basic types of boundary conditions: normal and tangential. 
For $\alpha\in\Omega^p(M)$, these can be stated in the form
$$
\eqalignno{
\hbox to 3 truecm{normal   :~~~\hfill}&
\iota_\partial{}^*\alpha=0;\hfill&(3.2.1a)\cr
\vphantom{\sum}
\hbox to 3 truecm{tangential:~~~\hfill}&
\iota_\partial{}^*j(n)\alpha=0.\hfill&(3.2.1b)\cr
}$$
It is not difficult to show that if $\alpha$ satisfies normal (tangential)
boundary conditions, then $\star\alpha$ satisfies tangential (normal)
boundary conditions. Note that $\alpha\in\Omega^p(M,\partial M)$ precisely when
$\alpha$ satisfies normal boundary conditions.

The basic choices of boundary conditions are: Dirichlet, Neumann, 
absolute and relative \ref{18,19}. For $\alpha\in\Omega^p(M)$, these take the 
form below.
$$
\eqalignno{\vphantom{\sum}
\hbox to 3 truecm{Dirichlet:~~~\hfill}&
\iota_\partial{}^*\alpha=0,
~~\iota_\partial{}^*j(n)\alpha=0;\hfill&(3.2.2a)\cr
\vphantom{\sum}
\hbox to 3 truecm{Neumann:~~~\hfill}&
\iota_\partial{}^*d^*\alpha=0,
~~\iota_\partial{}^*j(n)d\alpha=0;\hfill&(3.2.2b)\cr
\vphantom{\sum}
\hbox to 3 truecm{absolute:~~~\hfill}&
\iota_\partial{}^*j(n)\alpha=0,
~~\iota_\partial{}^*j(n)d\alpha=0;\hfill&(3.2.2c)\cr
\vphantom{\sum}
\hbox to 3 truecm{relative:~~~\hfill}&
\iota_\partial{}^*\alpha=0,
~~\iota_\partial{}^*d^*\alpha=0.\hfill&(3.2.2d)\cr
}
$$

For a set $\cal B$ of boundary conditions, we denote by $\Omega^p_{\cal B}(M)$ 
the space of $p$--forms obeying the boundary conditions $\cal B$. We further denote 
by $\Harm^p_{\cal B}(M)$ the corresponding space of harmonic $p$--forms 
and by $b_{{\cal B}p}$ its dimension ($p$--th $\cal B$ Betti number).

$\Omega^p(M)$ is a preHilbertian space with the standard inner product 
$\int_M\alpha\wedge\star\beta$, for $\alpha,~\beta\in\Omega^p(M)$.
When a set of boundary conditions $\cal B$ of the list (3.2.1) is imposed, 
the right hand side of (3.1.1) vanishes. So, (3.1.1) becomes a true
Hilbert space relation. (3.1.1) shows that $d^*$ is the adjoint of $d$ as 
suggested by the notation. Similarly, when a set of boundary conditions $\cal B$ 
of the list (3.2.2) is imposed, the right hand sides of the Green identities 
(3.1.2), (3.1.3) vanish. Again, (3.1.2), (3.1.3) become genuine Hilbert space 
relations. (3.1.2) implies that a $p$--form
$\alpha\in\Omega^p_{\cal B}(M)$ is harmonic if and only if it is closed an coclosed, 
i. e. $d\alpha=0$, $d^*\alpha=0$. (3.1.3) entails  
the Hodge Laplacian $\Delta$ on $\Omega^p_{\cal B}(M)$ 
is self adjoint.

For a form degree $p$ and a set $\cal B$ of boundary conditions of the list (3.2.2), 
we denote by $G_{\cal B}$ a $p$--form Green operator of the Hodge Laplacian $\Delta$ 
satisfying the boundary conditions $\cal B$ in both arguments (cf. eq. (3.1.4)). 
$G_{\cal B}$ exists and is unique only if the Betti number $b_{{\cal B}p}=0$.
 
There are several versions of Hodge orthogonal decomposition theorem on a 
manifold with boundary. Here, we report the ones which will be useful in the following.
$$
\eqalignno{\vphantom{\int}
\Omega^p(M)&=d\Omega^{p-1}_{\rm nor}(M)\oplus d^*\Omega^{p+1}_{\rm tan}(M)
\oplus \ker d\cap\ker d^*\cap\Omega^p(M),&(3.2.3a)\cr
\vphantom{\int}
\Omega^p(M)&=d\Omega^{p-1}(M)\oplus d^*\Omega^{p+1}_{\rm tan}(M)
\oplus \Harm^p_{\rm abs}(M),&(3.2.3b)\cr
\vphantom{\int}
\Omega^p(M)&=d\Omega^{p-1}_{\rm nor}(M)\oplus d^*\Omega^{p+1}(M)
\oplus \Harm^p_{\rm rel}(M).&(3.2.3c)\cr
}
$$

\vskip .3cm {\it  3.3. Harmonic representation of cohomology and Poincar\`e duality}

In the following, only absolute and relative boundary conditions 
will play a role.

On a manifold with boundary, both absolute and relative cohomology
have a harmonic representation. Indeed, one has the isomorphisms
$$
\eqalignno{
\vphantom{\int}
H^p(M,\Bbb R)&\cong\Harm^p_{\rm abs}(M), &(3.3.1a)\cr
\vphantom{\int}
H^p(M,\partial M,\Bbb R)&\cong\Harm^p_{\rm rel}(M),&(3.3.1b)\cr
}
$$
\ref{18,19}. 
(3.3.1) follows easily from the Hodge orthogonal decomposition
theorems (3.2.3b), (3.2.3c) and the equivalence of the de Rham cohomology
and the sheaf cohomology of $\Bbb R$.
This justifies the name given to these sets of boundary conditions. It also shows 
that the Betti numbers $b_{{\rm abs}p}$, $b_{{\rm rel}p}$ are finite.
The Hodge star operator $\star$ preserves harmonicity and interchanges absolute and 
relative boundary conditions. Thus, one has the isomorphism
$$
\matrix{
&{}_\star&\cr
\Harm^p_{\rm abs}(M)\!\!\!\!&\cong &\!\!\!\!\Harm^{m-p}_{\rm rel}(M).\cr
}
\eqno(3.3.2)
$$
As a consequence, 
$$
b_{{\rm abs}p}=b_{{\rm rel}m-p}.
\eqno(3.3.3)
$$
From (3.3.1), (3.3.2), one recovers the Poincar\'e duality relation
$$
H^p(M,\Bbb R)\cong H^{m-p}(M,\partial M,\Bbb R).
\eqno(3.3.4)
$$
The Poincar\'e duality pairing of forms $\alpha\in\Harm^p_{\rm abs}(M)$,
$\beta\in\Harm^{m-p}_{\rm rel}(M)$ is given as usual by $\int_M\alpha\wedge\beta$ 
\ref{18,19}.

We denote by $\Harm^p_{\rm abs\Bbb Z}(M)$, $\Harm^p_{\rm rel\Bbb Z}(M)$
the images in $\Harm^p_{\rm abs}(M)$, $\Harm^p_{\rm rel}(M)$
of the integer lattices $H_{\Bbb Z}^p(M,\Bbb R)$, $H_{\Bbb Z}^p(M,\partial M,
\Bbb R)$ of $H^p(M,\Bbb R)$, $H^p(M,\partial M,\Bbb R)$ under the isomorphisms 
(3.3.1), respectively. $\alpha\in\Harm^p_{\rm abs\Bbb Z}(M)$
if and only if $\int_S\alpha\in \Bbb Z$ for any singular $p$--cycle $S$ of $M$.
$\alpha\in\Harm^p_{\rm rel\Bbb Z}(M)$ if and only if 
$\int_S\alpha\in\Bbb Z$ for any relative singular $p$--cycle $S$ of $M$.

\vskip .3cm {\bf 4. The gauge theory action}
\vskip .3cm 

Let $M$ be a compact oriented $4$--fold with boundary $\partial M$. 
Let $g$ be a metric on $M$ and $g_\partial=\iota_\partial{}^*g$ be  
the induced metric on $\partial M$. 

\vskip .3cm {\it  4.1. The gauge theory action}

Let $P\in\Princ(M)$ and $P_\partial\in\Princ(\partial M)$ be principal
$\Bbb T$ bundles on $M$ and $\partial M$, respectively,  such that
$$
\iota_\partial{}^*P=P_\partial.
\eqno(4.1.1)
$$

The gauge theory action is the functional of the connection 
$A\in\Conn(P)$ defined by 
$$
S(A,\tau)={\sqrt{-1}\tau_2\over 4\pi}\int_M F_A\wedge \star F_A
+{\tau_1\over 4\pi}\int_MF_A\wedge F_A,
\eqno(4.1.2)
$$
where 
$$
\tau=\tau_1+\sqrt{-1}\tau_2,\quad 
\tau_1\in\Bbb R, \quad \tau_2\in\Bbb R_+
\eqno(4.1.3)
$$
(cf. eq. (2.4.1)).
$S(A,\tau)$ is invariant under the action of the group of gauge transformations
$\Gau(M)$ (cf. subsect. 2.3 and eq. (2.4.8)).

Since $M$ has a boundary, it is natural to require that the connection $A$
satisfies an appropriate set of boundary conditions. There is essentially only 
one natural choice of the latter, namely
$$
\iota_\partial{}^*A=A_\partial,
\eqno(4.1.4)
$$
where $A_\partial\in\Conn(P_\partial)$ is a fixed connection.
These boundary conditions are not preserved the action of $\Gau(M)$.
But they are by that of the group of relative gauge transformations 
$\Gau(M,\partial M)$ (cf. subsect. 2.3).

The field equations are obtained by varying $S(A,\tau)$ with respect to 
$A\in\Conn(P)$ with the boundary conditions (4.1.4) respected.
The allowed variations of $A$ are therefore $1$--forms 
$\delta A\in\Omega^1(M)$ such that $\iota_\partial{}^*\delta A=0$, i. e. 
$\delta A\in\Omega_{\rm nor}^1(M)$ satisfies normal boundary conditions 
(cf. eq. (3.2.1a)). Proceeding in this way, one finds that a gauge field 
$A$ satisfying the boundary conditions (4.1.4) is classical if it is solution of 
the usual vacuum Maxwell field equations
$$
d^*F_A=0.
\eqno(4.1.5)
$$

In view of quantizing the theory, we resort to the customary 
classical--background--quantum--splitting method, which we implement as follows.  
We factorize the bundle $P\in\Princ(M)$ as
$$
P=P_c\bar P,
\eqno(4.1.6)
$$
where $P_c\in\Princ(M)$ is a fiducial reference bundle such that
$$
\iota_\partial{}^*P_c=P_\partial
\eqno(4.1.7)
$$
and $\bar P\in\Princ(M)$ is a bundle such that $\iota_\partial{}^*\bar P$
is trivial. Next, we endow $\bar P$ with a trivialization 
$\bar t$ of $\iota_\partial{}^*\bar P$, obtaining in this way
a relative bundle $(\bar P,\bar t)\in\Princ(M,\partial M)$ (cf. subsect. 2.2). 
One needs $\bar t$ in order to carry out the decomposition 
of the connections of $P$ described below. 
Next, we choose fiducial reference connections $A_c\in\Conn(P)$ 
and $A_{c\partial}\in\Conn(P_\partial)$ such that 
$$
\iota_\partial{}^*A_c=A_{c\partial}.
\eqno(4.1.8)
$$
We further demand that $A_c$ satisfies the field equations (4.1.5),
$$
d^*F_{A_c}=0.
\eqno(4.1.9)
$$
We then write the generic connections $A\in\Conn(P)$ 
and $A_\partial\in\Conn(P_\partial)$ as follows:
$$
A=A_c+\bar A+{\cal A}+v,
\eqno(4.1.10)
$$
$$
A_\partial=A_{c\partial}+a.
\eqno(4.1.11)
$$
Here, $\bar A\in\Conn(\bar P,\bar t)$ is a relative connection and 
${\cal A}\in\Omega^1(M)$, $v\in\Omega^1(M)$ and $a\in\Omega^1(\partial M)$
are $1$--forms on $M$ and $\partial M$, respectively. 
$\bar A$ satisfies the boundary conditions (2.4.4),
$$
\iota_\partial{}^*F_{\bar A}=0
\eqno(4.1.12)
$$
and the field equations (4.1.5),
$$
d^*F_{\bar A}=0.
\eqno(4.1.13)
$$
By (2.4.2), (4.1.12), (4.1.13),
$F_{\bar A}\in\Harm_{\rm rel}^2(M)$ is a harmonic $2$--form 
satisfying relative boundary conditions (cf. eq. (3.2.2d)). 
As $F_{\bar A}$ is the curvature of a relative connection, $F_{\bar A}/2\pi$ has 
integer relative periods (cf. eq. (2.4.3)). 
Hence, $F_{\bar A}/2\pi\in\Harm_{{\rm rel}\Bbb Z}^2(M)$. 
${\cal A}$ satisfies the boundary conditions
$$
\iota_\partial{}^*{\cal A}=a,
\eqno(4.1.14)
$$
the field equations
$$
d^*d{\cal A}=0
\eqno(4.1.15)
$$
and the Lorenz gauge fixing condition
$$
d^*{\cal A}=0.
\eqno(4.1.16)
$$
$\cal A$ is determined by $a$ up to a certain ambiguity, 
as will be shown in the next subsection.
$v$ satisfies the normal boundary conditions
$$
\iota_\partial{}^*v=0.
\eqno(4.1.17)
$$
Thus, $v\in\Omega_{\rm nor}^1(M)$. 
$v$ is the bulk quantum fluctuation. Since this is a gauge theory, gauge fixing is 
required by quantization. We fix the gauge by imposing the customary Lorenz gauge 
fixing condition
$$
d^*v=0.
\eqno(4.1.18)
$$
Consequently, $v$ satisfies relative boundary conditions and, so, 
$v\in\Omega_{\rm rel}^1(M)$. It is easy to check that $A$ and $A_\partial$, 
as given in (4.1.10), (4.1.11), fulfill (4.1.4). 

A simple calculations shows that
$$
\eqalignno{
S(A,\tau)&=S(A_c+\bar A,\tau)+S({\cal A},\tau)
+{\sqrt{-1}\tau_2\over 2\pi}\oint_{\partial M}a\wedge
\star_\partial\iota_\partial{}^*j(n)F_{\bar A}
&(4.1.19)\cr
&+{\sqrt{-1}\tau_2\over 2\pi}\oint_{\partial M}a\wedge
\star_\partial\iota_\partial{}^*j(n)F_{A_c}+
+{\tau_1\over 2\pi}\oint_{\partial M}a\wedge
\iota_\partial{}^*F_{A_c}
+{\sqrt{-1}\tau_2\over 4\pi}\int_M dv\wedge \star dv.
&\cr
}
$$
To obtain (4.1.18), one exploits Stokes' theorem, the boundary conditions (4.1.12), 
(4.1.14) and (4.1.17) and the field equations (4.1.9), (4.1.13), (4.1.15).
The next step will be the computation of the various contributions in the 
right hand side of (4.1.19).

\vskip .3cm {\it  4.2. Calculation of the action $S({\cal A},\tau)$}

The calculation of the action $S({\cal A},\tau)$ requires the previous calculation 
of the $1$--form field ${\cal A}$. ${\cal A}$, in turn, should satisfy 
(4.1.14)--(4.1.16). Thus, the computation of ${\cal A}$ involves the solution 
of a certain boundary value problem. 
Abstractly, the problem can be stated in the following form  
$$
d^*d\omega=0,\quad d^*\omega=0,
\quad 
\eqno(4.2.1a)
$$
$$
\iota_\partial{}^*\omega=a,
\eqno(4.2.1b)
$$
with $\omega\in\Omega^1(M)$ and $\in\Omega^1(\partial M)$. 
Next, we shall discuss whether the problem admits 
a solution $\omega$ and whether the solution, when it exists, is unique.
Further, we shall provide an expression of $\omega$ in terms of $a$
valid under certain conditions.

In (4.2.1a), the condition $d^*\omega=0$ fixes the gauge. It is important to ascertain
whether it does so completely. To this end, let us consider the boundary value 
problem (4.2.1),  without gauge fixing:
$$
d^*d\omega=0,
\eqno(4.2.2a)
$$
$$
\iota_\partial{}^*\omega=a.
\eqno(4.2.2b)
$$
Assume that (4.2.2) has a solution $\omega$. If $\lambda\in\Omega^0(M)$ 
and $\iota_\partial{}^*d\lambda=d_\partial\iota_\partial{}^*\lambda=0$, 
then $\omega'=\omega+d\lambda$ is also a solution of (4.2.2). 
Of course, this degeneracy is due to the gauge symmetry of the problem (4.2.2).
Let us now impose on $\omega$ the gauge fixing condition $d^*\omega=0$.
Then, $d^*\omega'=0$, provided the gauge function $\lambda$ satisfies
$d^*d\lambda=0$. Since $d_\partial\iota_\partial{}^*\lambda=0$,
$\iota_\partial{}^*\lambda=b_01_\partial$ for some $b_0\in\Bbb R$. 
(Here, we assume $\partial M$ to be connected for simplicity.
The argument generalizes to the case where $\partial M$ has several connected 
components in straightforward fashion). Now define
$\bar\lambda=\lambda-b_01$. Then, $d^*d\bar\lambda=0$ and 
$\iota_\partial{}^*\bar\lambda=0$.
From the 1st Green identity (3.1.2) with $\alpha=\beta=\bar\lambda$, one finds that
$d\bar\lambda=0$. From the definition of $\bar \lambda$, it follows that 
$d\lambda=0$. Therefore, $\omega=\omega'$.
In conclusion, the gauge symmetry is completely fixed by the fixing condition. 

Let us analyze under which conditions the boundary value problem (4.2.1)
has a unique solution, assuming it has at least one. 
Let $\omega_1$, $\omega_2$ be two solutions of the problem
and let $\varpi=\omega_2-\omega_1$ be their difference. Then, $\varpi$ satisfies
$\Delta\varpi=0$, $\iota_\partial{}^*\varpi=0$, $\iota_\partial{}^*d^*\varpi=0$,
i. e. $\varpi\in\Harm_{\rm rel}^1(M)$ is a harmonic $1$--form obeying relative 
boundary conditions. Thus, from (3.3.1b), the solution of the boundary value 
problem (4.2.1), if any, is unique if the 1st relative cohomology space vanishes:
$$
H^1(M,\partial M,\Bbb R)=0.
\eqno(4.2.3)
$$

Let us assume first that (4.2.3) is fulfilled. 
As explained in subsect. 3.2, the vanishing of $b_{{\rm rel}1}$
ensures that there is a unique relative Green operator $G_{\rm rel}$ of the Hodge 
Laplacian $\Delta$ on $\Omega_{\rm rel}^1(M)$. $G_{\rm rel}$ satisfies
$$
\Delta G_{{\rm rel}x}=\star\delta_x,
\eqno(4.2.4a)
$$
$$
\iota_\partial{}^*G_{{\rm rel}x}=0,\quad \iota_\partial{}^*d^*G_{{\rm rel}x}=0,
\eqno(4.2.4b)
$$
for $x\in M\setminus\partial M$. 
Substituting $G_{\rm rel}$ in the identity (3.1.5) and using (4.2.4b), 
we conclude that, if the boundary value problem (4.2.1) has a solution $\omega$, 
this is necessarily given by the expression:
$$
\omega(x)=\oint_{\partial M}a\wedge
\star_\partial \iota_\partial{}^*j(n)d G_{{\rm rel}x},
\eqno(4.2.5)
$$
where $x\in M\setminus\partial M$. To show the existence of a solution of (4.2.1), 
we have only to verify that the right hand side of (4.2.5) satisfies (4.2.1). 

Recall that the spectrum of the Hodge Laplacian $\Delta$ on $\Omega_{\rm rel}^1(M)$
is discrete and that each eigenvalue has finite multiplicity \ref{18,19}. 
Let $\lambda_r$ be the eigenvalues of $\Delta$ on $\Omega_{\rm rel}^1(M)$
counting multiplicity and let $\phi_r$ be the corresponding normalized
$1$--eigenforms, so that $\Delta\phi_r=\lambda_r\phi_r$. 
Recalling that $b_{{\rm rel}1}=0$ and using the Hodge decomposition (3.2.3c), 
it is simple to show that $\lambda_r>0$ and that $\phi_r$ is either closed or 
coclosed, for every $r$. The relative Green function is therefore 
$$
G_{\rm rel}(x,x')=-\sum_r \lambda_r{}^{-1}\phi_r(x)\phi_r(x'),
\eqno(4.2.6)
$$ 
for $x,~x'\in M$. In (4.2.5), there appears the kernel 
$$
H(x,x')=(\iota_\partial{}^*j(n)d)(x')G_{\rm rel}(x,x'),
\eqno(4.2.7)
$$
where $x\in M\setminus\partial M$, $x'\in\partial M$. 
Substituting (4.2.6) in (4.2.7), we find that 
$$
H(x,x')=-\sum_r \lambda_r{}^{-1}\phi_r(x)(\iota_\partial{}^*j(n)d)(x')\phi_r(x').
\eqno(4.2.8)
$$  
Using (4.2.8), it is straightforward to show that 
$$
\Delta(x)H(x,x')=0, \quad d^*(x)H(x,x')=0,
\eqno(4.2.9)
$$
for $x\in M\setminus\partial M$, $x'\in\partial M$.
To this end, we note that $\sum_r\phi_r(x)\star\phi_r(x')=\delta_x(x')=0$,
for $x\in M\setminus\partial M$, $x'\in\partial M$ and that, 
in (4.2.8), only the coclosed $\phi_r$ contribute.
From (4.2.9), it is now apparent that $\omega$ as given by (4.2.5)
satisfies (4.2.1a) in $M\setminus\partial M$. 
It remains to check that $\omega$ extends smoothly to $\partial M$ and that 
the boundary condition (4.2.1b) is verified. We do not know how to do this in 
general. Below, we assume that the verification is possible.

When (4.2.3) does not hold, the above discussion must be modified. There is no reason 
in principle why the problem (4.2.1) should not have a solution $\omega$. In this
case, however, $\omega$ is determined by $a$ only up to the addition of an arbitrary 
element $\varpi\in\Harm_{\rm rel}^1(M)$ and is no longer given by a simple expression 
of the form (4.2.5). The combination $d\omega$, conversely,  is uniquely 
determined by $a$, since $d\varpi=0$. From (4.2.1), it is obvious that 
$d\omega$ depends linearly on $a$.

Let us assume again that (4.2.3) is fulfilled.
From (4.1.14)--(4.1.16), ${\cal A}$ satisfies the boundary value problem (4.2.1). 
As (4.2.3) holds, it is given by (4.2.5).
Using the above considerations, it is now easy to compute the action 
$S({\cal A},\tau)$. One finds
$$
S({\cal A},\tau)={\sqrt{-1}\tau_2\over 4\pi}\oint _{\partial M}
a\wedge\star_\partial \iota_\partial{}^*j(n)d {\cal A}
+{\tau_1\over 4\pi}\oint _{\partial M}a\wedge d_\partial a,
\eqno(4.2.10)
$$
where, in the first term,
$$
{\cal A}(x)=\oint_{\partial M}a\wedge
\star_\partial\iota_\partial{}^*j(n)d G_{{\rm rel}x}. 
\eqno(4.2.11)
$$
Note that the first term contains effectively the kernel 
$$
K(x,x')=(\iota_\partial{}^*j(n)d)(x)(\iota_\partial{}^*j(n)d)(x')G_{\rm rel}(x;x'),
\eqno(4.2.12)
$$
where $x,~x'\in\partial M$. $K$ is a symmetric distributional 
biform of bidegree $(1,1)$ on $\partial M\times\partial M$.

It is convenient to write the above expression in more explicit tensor notation.
For every point $x\in\partial M$, there is an open neighborhood 
$O$ of $x$ in $M$ and a diffeomorphism 
$\phi:O\rightarrow(\Bbb R_-\cup\{0\})\times \Bbb R^3$ such that 
$\phi(O\cap\partial M)=\{0\}\times\Bbb R^3$. So, there are local coordinates $x^i$, 
$i=1,~2,~3,~4$, at the boundary, which split as $(x^0,x^a)$, where $x^0$ is valued 
in $\Bbb R_-\cup\{0\}$ and $x^a$ is valued in $\Bbb R$, $a=1,~2,~3$. Locally,
the boundary is defined by the condition $x^0=0$ and is parameterized by the
$x^a$. We denote by middle lower case Latin letters $i,~j,~k$, ... bulk 4 
dimensional tensor indices and by early lower case Latin letters $a,~b,~c$, ... 
boundary 3 dimensional tensor indices. The outward unit normal vector field is 
given by 
$$
n^i=g^{0i}/g^{00}{}^{1\over 2}.
\eqno(4.2.13)
$$
The induced metric $g_\partial$ on $\partial M$ and the totally antisymmetric tensor
$\epsilon_\partial$ are given by
$$
g_{\partial ab}=g_{ab}\circ\iota_\partial, \quad
g_\partial{}^{ab}=\big(g^{ab}-g^{a0}g^{0b}/g^{00}\big)\circ\iota_\partial,
\eqno(4.2.14)
$$
$$
\epsilon_{\partial abc}
=\big(g^{00}{}^{1\over 2}\epsilon_{0abc}\big)\circ\iota_\partial.
\eqno(4.2.15)
$$
The action $S({\cal A},\tau)$ reads then 
$$
\eqalignno{
S({\cal A},\tau)
&={\sqrt{-1}\tau_2\over 4\pi}\oint _{\partial M}\oint _{\partial M}
d^3xg_\partial{}^{1\over 2}(x)d^3x'g_\partial{}^{1\over 2}(x')
g_\partial{}^{ab}(x)g_\partial{}^{a'b'}(x')&(4.2.16)\cr
&\phantom{=}\times a_b(x)a_{b'}(x')K_{aa'}(x;x')
+{\tau_1\over 4\pi}\oint _{\partial M}d^3xg_\partial{}^{1\over 2}(x)
\epsilon_\partial{}^{abc}(x)a_a(x)\partial_ba_c(x),&\cr
}
$$
where the kernel $K_{aa'}(x;x')$ is given by 
$$
\eqalignno{
\vphantom{1\over 2}
K_{aa'}(x;x')&=g^{00}{}^{1\over 2}(x)g^{0'0'}{}^{1\over 2}(x')
\big[\partial_0\partial'{}_{0'}G_{{\rm rel}aa'}(x,x')
-\partial_0\partial'{}_{a'}G_{{\rm rel}a0'}(x,x')&(4.2.17)\cr
\vphantom{1\over 2}&\phantom{g^{00}{}^{1\over 2}(x)g^{0'0'}{}^{1\over 2}(x')\big[}
-\partial_a\partial'{}_{0'}G_{{\rm rel}0a'}(x,x')
+\partial_a\partial'{}_{a'}G_{{\rm rel}00'}(x,x')\big],&\cr
}
$$
with $x,x'\in\partial M$.

When (4.2.3) does not hold, the above calculation is not applicable.
For reasons explained above, even though $\cal A$ is not uniquely determined
by $a$, being affected by an additive ambiguity in $\Harm_{\rm rel}^1(M)$,
the combination $d{\cal A}$ is and depends linearly on $a$.
Since the action $S({\cal A},\tau)$ depends quadratically on 
$d{\cal A}$, by (4.1.2), $S({\cal A},\tau)$ is a well defined 
quadratic functional of $a$. It is easy to see that
$S({\cal A},\tau)$ is still given by an expression of the form (4.2.16).
However, now, the kernel $K$ in no longer given by (4.2.17).

So, regardless whether (4.2.3) holds or not, we can write in general
$$
S({\cal A},\tau)=S(a,\tau),
\eqno(4.2.18)
$$
where $S(a,\tau)$ is the quadratic functional of $a$ given by the right hand side of
(4.2.16) in terms of the kernel $K$. When (4.2.3) is fulfilled, $K$ is given by 
(4.2.17).

\vskip .3cm {\it  4.3. Calculation of the action $S(A_c+\bar A,\tau)$}

Let $\{F_r|r=1,\ldots,b_2\}$ be a basis of the integer lattice 
$\Harm^2_{{\rm rel}\Bbb Z}(M)$ of harmonic 2 forms obeying 
relative boundary conditions and having relative integer periods,
where $b_2\equiv b_{{\rm rel}2}=b_{{\rm abs}2}$ (cf. eq. (3.2.2d) 
and subsects. 3.2, 3.3). 

The intersection matrix $Q$ of $M$ is defined by 
$$
Q_{rs}=\int_M F_r\wedge F_s.
\eqno(4.3.1)
$$
As well known, $Q$ is a symmetric integer $b_2\times b_2$ matrix
characterizing the topology of $M$. When $\partial M\not=\emptyset$,
$Q$ is generally singular. (Recall that, instead, if $\partial M=\emptyset$, 
$Q$ is unimodular).

The Hodge matrix $H$ of $M$ is defined by 
$$
H_{rs}=\int_M F_r\wedge \star F_s.
\eqno(4.3.2)
$$
As $\int_M F\wedge *F$ is a norm on $\Harm^2(M)$,  
$H$ is a positive definite symmetric $b_2\times b_2$ matrix.

From (4.1.12), (4.1.13) and the remarks below those relations
one has immediately that
$$ 
F_{\bar A}=2\pi\sum_rk^rF_r,
\eqno(4.3.3)
$$
where $k^r\in \Bbb Z$. So, $F_{\bar A}$ is completely characterized by the 
lattice point $k\in\Bbb Z^{b_2}$.

Using (4.1.2), (4.3.1)--(4.3.3), we obtain
$$
\eqalignno{
S(A_c+\bar A,\tau)&=S(A_c,\tau)+\pi k^t(\tau_1Q+\sqrt{-1}\tau_2H)k
\vphantom{\int_M}
&(4.3.4)\cr
&+k^t\bigg[\sqrt{-1}\tau_2\int_M F\wedge \star F_{A_c}
+\tau_1\int_M F\wedge F_{A_c}\bigg],
&\cr
}
$$
where matrix notation is used.

\vskip .3cm {\it  4.4. Calculation of the term ${\sqrt{-1}\tau_2\over 2\pi}
\oint_{\partial M}a\wedge\star_\partial\iota_\partial{}^*j(n)F_{\bar A}$}

Using (4.3.3), this term is easily computed. The result is 
$$
{\sqrt{-1}\tau_2\over 2\pi}\oint_{\partial M}a\wedge
\star_\partial\iota_\partial{}^*j(n)F_{\bar A}
=\sqrt{-1}\tau_2 k^t\oint_{\partial M}a\wedge
\star_\partial \iota_\partial{}^*j(n)F,
\eqno(4.4.1)
$$
where matrix notation is used again.

The computation of $S(A,\tau)$ is now complete.

\vskip .3cm {\bf 5. $4$--d Abelian duality and $\SL(2,\Bbb Z)$ action 
\bf of $3$--d conformal field theory}

\vskip .3cm 

We now have all the results needed to discuss Witten's claim in \ref{1} that 
the $\SL(2,\Bbb Z)$ action in  nearly Gaussian $3$--dimensional conformal field 
theory is a holographic image of the well--known Abelian duality of $4$--dimensional 
gauge theory. \footnote{}{}\footnote{${}^2$}
{I thank E. Witten for explaining me some crucial points of the discussion 
of subsects. 5.1, 5.2 below.}

\vskip .3cm {\it  5.1. The $\SL(2,\Bbb Z)$ action}

Let $\hat M$ be an oriented compact $3$--fold. Let $\hat M$ be endowed with a 
conformal structure $\hat\gamma$ and a compatible spin structure $\hat\sigma$. 
(Below, we shall denote all objects relating to $\hat M$ with a hat). 
Consider a $3$--dimensional conformal field theory on $\hat M$ having a global 
$\U(1)$ symmetry with symmetry current $\hat\jmath$. 
Fix a bundle $\hat P\in\Princ(\hat M)$ and couple $\hat\jmath$ to a background 
gauge connection $\hat A\in\Conn(\hat P)$ (cf. subsect. 2.2, 2.4). 
The coupling yields a generating functional $Z_{\hat P}(\hat A)$ 
of the correlators of $\hat\jmath$ depending on the bundle $\hat P$
of a form analogous to (1.1).

As explained in the introduction, the functional $Z_{\hat P}(\hat A)$ completely 
characterizes the conformal field theory with the chosen gauge coupling. 
So, the $\SL(2,\Bbb Z)$ action on the set of all such field theories with 
gauge coupling can be expressed as the action of $\SL(2,\Bbb Z)$ operators 
$\widehat S$, $\widehat T$, $\widehat C$ on the family of the corresponding 
functionals $Z_{\hat P}(\hat A)$. 

Taking (1.5) as a model, the expression of the operators $\widehat S$, 
$\widehat T$, $\widehat C$ can be cast in a relatively more precise fashion as 
follows:
$$
\eqalignno{
\vphantom{\int}
\widehat S Z_{\hat P}(\hat A)
&=\sum_{\hat Q\in\Princ(\hat M)}\,\,
\int\limits_{\hat B\in\Conn(\hat Q)} {D\hat B\over\varrho(\hat M)} 
Z_{\hat Q}(\hat B)
\exp\bigg({\sqrt{-1}\over 2\pi}\int_{\hat M}\hat B\wedge \hat d\hat A\bigg),
&(5.1.1a)\cr
\vphantom{\int\limits_{\hat B\in\Conn(\hat Q)}}
\widehat T Z_{\hat P}(\hat A)
&=Z_{\hat P}(\hat A)
\exp\bigg({\sqrt{-1}\over 4\pi}\int_{\hat M}\hat A\wedge \hat d \hat A\bigg),
&(5.1.1b)\cr
\vphantom{\int}
\widehat C Z_{\hat P}(\hat A)
&=Z_{\hat P{}^{-1}}(-\hat A),
&(5.1.1c)\cr
}
$$
where $\varrho(\hat M)={\rm vol}(\Gau(\hat M))$ is the formal volume of the group
of gauge transformations on $\hat M$ (cf. subsect. 2.3). 
The level $1$ mixed Chern Simons action
${1\over 2\pi}\int_{\hat M}\hat B\wedge\hat d\hat A$ with $\hat A\in\Conn(\hat P)$, 
$\hat B\in\Conn(\hat Q)$ appearing in (5.1.1a) is defined according to the procedure 
expounded in Witten's paper \ref{1}.
This requires choosing an oriented compact $4$--fold $M$ with $\partial M=\hat M$ 
and bundles $P,~Q\in\Princ(M)$ extending $\hat P,~\hat Q$
together with connections $A\in\Conn(P)$, $B\in\Conn(Q)$ extending $\hat A,~\hat B$.
Then, ${1\over 2\pi}\int_{\hat M}\hat B\wedge\hat d\hat A
={1\over 2\pi}\int_M F_B\wedge F_A$ mod $2\pi \Bbb Z$ independently from all choices. 
Recall that the extendability of a bundle $\hat R\in\Princ(\hat M)$ to a bundle 
$R\in\Princ(M)$ is governed by the exact sequence (2.5.1). In general, for a given 
$3$--fold $\hat M$, there is no universal choice of $M$ doing the job.
Several $M$ may be needed to allow for the extension of all possible bundles
$\hat R\in\Princ(\hat M)$. In particular, for a pair of bundles 
$\hat P,~\hat Q\in\Princ(\hat M)$, there may be no $4$--fold $M$ 
allowing for the simultaneous extension of both. In that case, 
the corresponding term in the sum over $\hat Q\in\Princ(\hat M)$ in (5.1.1a) is 
supposed to be absent. The level $1/2$ Chern Simons action
${1\over 4\pi}\int_{\hat M}\hat A\wedge \hat d \hat A$ with $\hat A\in\Conn(\hat P)$
appearing in (5.1.1b) is defined again according to the procedure expounded in 
Witten's paper. In this case, it is necessary to choose an oriented 
compact $4$--fold $M$ with $\partial M=\hat M$, such that there are a metric $g$ 
of $M$ whose pull--back to $\hat M$ belongs to $\hat\gamma$ and a spin structure 
$\sigma$ of $M$ subordinated to $g$ extending $\hat\sigma$, and a bundle 
$P\in\Princ(M)$ extending $\hat P$ together with a connection $A\in\Conn(P)$ 
extending $\hat A$. Then, ${1\over 4\pi}\int_{\hat M}\hat A\wedge\hat d\hat A
={1\over 4\pi}\int_M F_A\wedge F_A$ mod $2\pi \Bbb Z$ independently from all choices. 
Again, for reasons already explained, there is no universal choice of $M$
doing the job in general.

Next, let us analyze the gauge invariance of the functional $Z_{\hat P}(\hat A)$. 
According to (1.2), for given bundle $\hat P\in\Princ(\hat M)$ 
and connection $\hat A\in\Conn(\hat P)$, $Z_{\hat P}(\hat A)$ should 
be invariant under any topologically trivial gauge transformation:
$$
Z_{\hat P}(\hat A^{\hat U})=Z_{\hat P}(\hat A),
\eqno(5.1.2)
$$
for $\hat U\in\Gau_c(\hat M)$ (cf. subsect. 2.3 and eq. (2.4.8)).
More generally, one may demand that $Z_{\hat P}(\hat A)$ should 
be invariant under any gauge transformation $\hat U\in\Gau(\hat M)$.
However, there are problems associated with this point of view.
If we insist that $Z_{\hat P}(\hat A)$ is the generating functional of the correlators 
of the conserved current $\hat\jmath$ and, thus, is of a form analogous to 
(1.1), (5.1.2) must definitely hold for $\hat U\in\Gau_c(\hat M)$. 
Conversely, (5.1.2) can hold for a general $\hat U\in\Gau(\hat M)$ only if 
the current $\hat\jmath$ satisfies in addition a suitable quantization condition, 
which is not guaranteed in general. (Recall that, by (2.4.8),
$\hat A^{\hat U}=\hat A+\hat B_{\hat U}$, where 
$\hat B_{\hat U} \in\Omega_{\Bbb Z}^1(\hat M)$ is a closed $1$--form with 
integer periods). 
Further, the $\widehat S$, $\widehat T$ operators defined in (5.1.1a), (5.1.1b) 
are manifestly compatible with (5.1.2) if $\hat U\in\Gau_c(\hat M)$,
but they are not if $\hat U\in\Gau(\hat M)$. In fact, 
the gauge transformation of the Chern Simons actions 
entering (5.1.1a), (5.1.1a) requires extending a gauge 
transformation $\hat U\in\Gau(\hat M)$ to a gauge transformation
$U\in\Gau(M)$, where $M$ is the compact oriented $4$--fold with 
$\partial M=\hat M$ required by the definition of those actions.
Recall that the extendability of a gauge transformation 
$\hat V\in\Gau(\hat M)$ to a gauge transformation $V\in\Gau(M)$ is governed by 
the exact sequence (2.5.5) and, for a given $M$, is not guaranteed in general. 
So, it seems plausible that one should demand only the restricted version 
of gauge invariance (5.1.2). Besides, $\widehat S Z_{\hat P}(\hat A)$, 
as defined in (5.1.1a), has obviously a larger invariance: 
it is invariant under shifts of $\hat A$ by a closed $1$--form 
$\hat B\in\Omega^1(\hat M)$. This does not seem to be
compatible with the interpretation of 
$\widehat S Z_{\hat P}(\hat A)$ as generating functional of the correlators of a
conserved current $\hat\jmath$ such as (1.1), unless all closed $1$--forms 
$\hat B\in\Omega^1(\hat M)$ are exact, i. e. 
$$ 
H^1(\hat M,\Bbb R)=0.
\eqno(5.1.3)
$$
Below, we shall therefore consider only $3$-folds $\hat M$ satisfying (5.1.3).
Since $\hat M$ is a compact oriented $3$--fold, $\Tor H^1(\hat M,\Bbb Z)=0$.
(This is a simple consequence of the universal coefficient theorem for cohomology).
As a consequence, one has 
$$ 
H^1(\hat M,\Bbb Z)=0.
\eqno(5.1.4)
$$
Then, the short exact sequence (2.3.3) entails that 
$$ 
\Gau(\hat M)=\Gau_c(\hat M).
\eqno(5.1.5)
$$
Hence, all gauge transformations on $\hat M$ are topologically trivial.

By Poincar\`e duality, (5.1.3) implies that
$$ 
H^2(\hat M,\Bbb R)=0.
\eqno(5.1.6)
$$
It follows that 
$$ 
H^2(\hat M,\Bbb Z)=\Tor H^2(\hat M,\Bbb Z).
\eqno(5.1.7)
$$
By the isomorphism (2.2.2), we have then
$$ 
\Princ(\hat M)=\Princ_0(\hat M).
\eqno(5.1.8)
$$
So, all principal bundles on $\hat M$ are flat.

Before proceeding, one should keep in mind, however, that (5.1.3) is a mere 
simplifying assumption, not a consistency requirement. (5.13) may indeed be relaxed, 
though  at the price of complicating the formalism.

\vskip .3cm {\it  5.2. Witten's holographic conjecture}

Let us now come to Witten's conjecture. For the class of large $N$ nearly 
Gaussian $3$ dimensional conformal field theories on an oriented 
compact $3$--fold $\hat M$ with conformal structure $\hat\gamma$ and compatible spin 
structure $\hat\sigma$, considered by Witten, the generating functional 
$Z_{\hat P}(\hat A,\tau)$ depends on a modular parameter $\tau\in{\Bbb H}_+$, 
parameterizing the various field theories, and satisfies the modular relations 
$$
\eqalignno{
\vphantom{\int}
\widehat S Z_{\hat P}(\hat A,\tau)&=Z_{\hat P}(\hat A,-1/\tau),
&(5.2.1a)\cr
\vphantom{\int}
\widehat T Z_{\hat P}(\hat A,\tau)&=Z_{\hat P}(\hat A,\tau+1),
&(5.2.1b)\cr
\vphantom{\int}
\widehat C Z_{\hat P}(\hat A,\tau)
&=Z_{\hat P}(\hat A,\tau),
&(5.2.1c)\cr
}
$$
with $\widehat S$, $\widehat T$, $\widehat C$ defined by (5.1.1) (cf. eq. (1.7)).
According to Witten's holographic conjecture, these $3$ dimensional conformal 
field theories are obtainable via holography from $4$ dimensional Abelian gauge theory 
on $4$--folds $M$ bounded by $\hat M$. The $\tau$ dependence of 
$Z_{\hat P}(\hat A,\tau)$ is a consequence of that of the gauge theory action 
$S(A,\tau)$ (cf. eqs. (4.1.2), (4.1.3)). Finally, the relations (5.2.1) follow from 
the Abelian duality of gauge theory \ref{11--15}.

The conjecture can be formulated in a relatively more precise fashion as follows. 
In the holographic correspondence, the conformal field theory on the given 
$3$--fold $\hat M$ is constructed by summing over $4$--folds $M$ that have $\hat M$ 
as their boundary. In general, no particular $M$ contributes the full answer and 
the individual $M$, which do contribute, do so only to some topological sectors of the 
conformal field theory \ref{20,21}. The conjecture can thus be stated as
$$
Z_{\hat P}(\hat A,\tau)=\sum_{M\in{\eul M}(\hat P)}Z^M_{\hat P}(\hat A,\tau),
\eqno(5.2.2)
$$
where ${\eul M}(\hat P)$ is an appropriate set of oriented compact  
$4$--folds $M$ such that $\partial M=\hat M$ and $Z^M_{\hat P}(\hat A,\tau)$ is 
the partition function of Abelian gauge theory on $M$ with boundary conditions 
specified by the data $\hat P$, $\hat A$. Unfortunately, neither the content of the 
set ${\eul M}(\hat P)$ nor the relative normalization of the 
individual contributions $Z^M_{\hat P}(\hat A,\tau)$ are known in general.

For a given $\hat P\in\Princ(\hat M)$, each $4$--fold $M\in{\eul M}(\hat P)$ must have 
the following properties. First, in compliance with the principles of the 
AdS/CFT correspondence, there exists an asymptotically hyperbolic Einstein metric 
$g_{AHE}$ of $M$ with a double pole at $\hat M$ having the conformal structure 
$\hat\gamma$ of $\hat M$ as conformal infinity (see ref. \ref{22}). 
Second, there exists a spin structure $\sigma$ of $M$ subordinated to some 
conformal compactification $g$ of $g_{AHE}$ extending the spin structure 
$\hat\sigma$ of $\hat M$ (see again ref. \ref{22}). 
Third, there exists a bundle $P\in\Princ(M)$ extending $\hat P$. 
\footnote{}{}\footnote{${}^3$}
{In reference \ref{20}, it is argued that, in general, there might be additional 
contributions from branes on $M$. For these, this condition may be relaxed.
For simplicity, we shall ignore this possibility in the following analysis.}
Finally, the set ${\eul M}(\hat P)$ must be independent from the data 
$\hat\gamma$, $\hat \sigma$ at least for $\hat\gamma$, $\hat\sigma$  
belonging to a reasonably broad class. 

For given $\hat \sigma$, $\hat\gamma$, $\hat P\in\Princ(\hat M)$ and 
$M\in{\eul M}(\hat P)$, an explicit expression of $Z^M_{\hat P}(\hat A,\tau)$ 
can be obtained. To be consistent with the notational conventions of sect. 4, 
we set $\hat\sigma=\sigma_\partial$, $\hat\gamma=\gamma_\partial$, 
$\hat P=P_\partial$ and, accordingly, we denote a generic connection 
$\hat A\in\Conn(\hat P)$ as $A_\partial\in\Conn(P_\partial)$
and the functional $Z^M_{\hat P}(\hat A,\tau)$ as $Z^M_{P_\partial}(A_\partial,\tau)$. 
The gauge theory partition function $Z^M_{P_\partial}(A_\partial,\tau)$
involves a sum over the set of the asymptotically hyperbolic Einstein metrics 
$g_{AHE}$ of $M$ with a double pole at $\partial M$ having $\gamma_\partial$ as 
conformal infinity modulo the action of the diffeomorphisms of $M$ inducing the 
identity on $\partial M$ and, for each such $g_{AHE}$, a sum over the set of the 
spin structures $\sigma$ of $M$ subordinated to some conformal compactification $g$ 
of $g_{AHE}$ and extending $\sigma_\partial$. It further involves a sum over the set 
of the bundles $P\in\Princ(M)$ such that $\iota_\partial{}^*P=P_\partial$ and a 
functional integration over the set of the connections $A\in\Conn(P)$ such that 
$\iota_\partial{}^*A=A_\partial$ \ref{11--15}. 
However, below, in order to keep the formulas reasonably simple, we shall not 
explicitly indicate the first two summations. Taking all this into account, 
$Z^M_{P_\partial}(A_\partial,\tau)$ is given by 
$$
Z^M_{P_\partial}(A_\partial,\tau)
=\sum_{P\in\Princ(M),\iota_\partial{}^*P=P_\partial}\,\,
\int\limits_{A\in\Conn(P),\iota_\partial{}^*A=A_\partial}
{DA\over\varrho(M,\partial M)}\exp\left(\sqrt{-1}S(A,\tau)\right),
\eqno(5.2.3)
$$
where $S(A,\tau)$ is the gauge theory action (4.1.2) and 
$\varrho(M,\partial M)={\rm vol}(\Gau(M,\partial M))$ is the formal volume 
of the group of relative gauge transformations (cf. subsect. 2.3).
The metric $g$ contained in the action $S(A,\tau)$ is a conformal 
compactification of the relevant asymptotically hyperbolic Einstein metric 
$g_{AHE}$.

Following the treatment of sect. 4, we decompose the bundle $P$ as in (4.1.6)
and the connections $A$ and $A_\partial$ as in (4.1.10), (4.1.11).
Proceeding in this way, (5.2.3) can be cast as
$$
Z^M_{P_\partial}(A_{c\partial}+a,\tau)=
\sum_{(\bar P,\bar t)\in\Princ(M,\partial M)}\,\,
\int\limits_{v\in\Omega^1_{\rm nor}}{Dv\over\varrho(M,\partial M)}
\exp\left(\sqrt{-1}S(A_c+\bar A+{\cal A}+v,\tau)\right).
\eqno(5.2.4) 
$$
Recall that, here, $\Princ(M,\partial M)$ is the group of relative principal bundles
on $M$ and $\bar A\in\Conn(\bar P,\bar t)$ is a chosen relative connection of the 
relative bundle $(\bar P,\bar t)$ (cf. subsect. 2.2, 2.4). The action $S(A,\tau)$ 
in the above expression is given by (4.1.19), (4.2.18), (4.3.4) and (4.4.1).
The functional integration of the quantum fluctuations $v$ is Gaussian and 
thus trivial. It yields a factor $\tau_2^{b_{{\rm rel}1}/2}$ for reasons 
analogous to those leading to a factor $\tau_2^{(b_1-1)/2}$ in the boundaryless
case of \ref{12,15}. On account of (2.2.4), the sum over $\Princ(M,\partial M)$
can be turned into a sum over the degree $2$ relative $\Bbb Z$ cohomology group
$H^2(M,\partial M,\Bbb Z)$. The integrand depends only on the free part 
of the latter, as is evident from the calculation of subsect. 4.3.
So, the sum over $H^2(M,\partial M,\Bbb Z)$ reduces to one over the lattice 
$\Bbb Z^{b_2}$ times a factor $t_{\rm rel}^2=|\Tor H^2(M,\partial M,\Bbb Z)|$.
Therefore, the $\tau$ dependent factor of $Z^M_{P_\partial}(A_{c\partial}+a,\tau)$ 
is of the form 
$$
\eqalignno{
Z^M_{P_\partial}&(A_{c\partial}+a,\tau)=
\exp\bigg\{\sqrt{-1}\bigg[S(A_c,\tau)+S(a,\tau)
+{\sqrt{-1}\tau_2\over 2\pi}\oint_{\partial M}a\wedge
\star_\partial\iota_\partial{}^*j(n)F_{A_c}~~~~~~~~&(5.2.5)\cr
&+{\tau_1\over 2\pi}\oint_{\partial M}a\wedge
\iota_\partial{}^*F_{A_c}\bigg]\bigg\}\cdot t_{\rm rel}^2 \,\tau_2^{b_{{\rm rel}1}/2}
\sum_{k\in\Bbb Z^{b_2}}\exp\bigg\{\sqrt{-1}\bigg[
\pi k^t(\tau_1Q+\sqrt{-1}\tau_2H)k&\cr
&+k^t\bigg(\sqrt{-1}\tau_2\int_M F\wedge \star F_{A_c}
+\tau_1\int_M F\wedge F_{A_c}
+\sqrt{-1}\tau_2\oint_{\partial M}a\wedge\star_\partial\iota_\partial{}^*j(n)F
\bigg)\bigg]\bigg\}.
&\cr
}
$$

To convince ourselves that Witten's conjecture is reasonable, let us examine a fully 
computable example in detail. 
Suppose that $\hat M=S^3$ and that $\hat\gamma$ and $\hat\sigma$
are respectively the conformal structure of the standard round metric of $S^3$ and 
the spin structure of $S^3$ subordinated to it. ($\hat\sigma$ is uniquely determined
for $H^1(S^3,\Bbb Z_2)=0$). Since $H^1(S^3,\Bbb Z)=0$, (5.1.3) is satisfied. 
As $H^2(S^3,\Bbb Z)=0$, $\Princ(S^3)$ contains only the trivial bundle $\hat 1$, 
by the isomorphism (2.2.2). 
There is a distinguished $4$--fold $M$ with $\partial M=S^3$, namely $M=B^4$. 
$B^4$ admits an asymptotically hyperbolic Einstein metric with a double pole 
at $\partial B^4$ having the conformal structure $\hat\gamma$ 
as conformal infinity, namely the standard Poincar\'e metric $g_P$. The latter is 
also the only such metric modulo diffeomorphisms of $B^4$ inducing the identity on 
$\partial B^4$. Further, $B^4$ supports a unique spin structure $\sigma$
subordinated to the obvious conformal compactification $g$ of $g_P$, 
which necessarily extends the spin structure $\hat\sigma$.
Bundle extendability is obviously not an issue in this case. 
It is therefore reasonable to assume, given the simplicity of the topological setting, 
that ${\eul M}(\hat 1)=\{B^4\}$. (5.2.2) then states that 
$$
Z_{\hat 1}(\hat A,\tau)=Z^{B^4}_{\hat 1}(\hat A,\tau).
\eqno(5.2.6)
$$

Choosing conveniently $A_{c\partial}=0$, $P_c=1$ and $A_c=0$ and recalling that
$B^4$ is cohomologically trivial, (5.2.5) yields simply
$$
Z^{B^4}_{\hat 1}(a,\tau)=\exp\big(\sqrt{-1}S(a,\tau)\big).
\eqno(5.2.7)
$$ 
As conformal compactification of $g_P$, we take the flat metric $g$ induced by 
the natural inclusion of $B^4$ into $\Bbb R^4$ with standard flat metric. 
Since $B^4$ satisfies condition (4.2.3), the calculation of 
$S(a,\tau)$ of subsect. 4.2 can be used. 

Let $O$ be the open subset of $B^4$ obtained by removing a closed segment joining 
the center of $B^4$ and a point of the boundary $\partial B^4$. 
Using appropriate adapted local coordinates $x^i=(x^0,\ul x)$ as in subsect. 4.2, 
we can identify $O$ with $(\Bbb R_-\cup\{0\})\times \Bbb R^3$ and $g_{ij}$ with 
$\delta_{ij}$. When the support of $a$ is contained in $O$, as we assume
below, the computation is amenable by standard calculus.

The relative Green operator $G_{{\rm rel}ij'}(x;x')$ on $1$--forms
(cf. subsect. 4.2) is given explicitly by  
$$
G_{{\rm rel}ij'}(x;x')=
-{1\over 4\pi^2}\Big[{\delta_{ij'}\over (x^0-x'^{0'})^2+(\ul x-\ul x')^2}
-{\delta_{ij'}-2\delta_{i0}\delta_{0'j'}
\over(x^0+x'^{0'})^2+(\ul x-\ul x')^2}\Big].
\eqno(5.2.8) 
$$
The first term is harmonic for $x'\not=x$ and ensures that the 
distributional equation 
$$
\partial^k\partial_kG_{{\rm rel}ij'}(x;x')
=\delta_{ij'}\delta(x-x')
\eqno(5.2.9) 
$$
is satisfied, as required by (4.2.4a). 
The second term is harmonic everywhere and is required to make sure 
that the relative boundary conditions 
$$
G_{{\rm rel}ia'}(x;0,\ul x')=0, 
\quad \partial^{\prime j'}G_{{\rm rel}ij'}(x;0,\ul x')=0
\eqno(5.2.10)
$$
are fulfilled, in compliance with (4.2.4b).

From (4.2.17), the kernel $K_{aa'}(\ul x;\ul x')$ is given by
$$
K_{aa'}(\ul x;\ul x')=
\partial_0\partial'{}_{0'}G_{{\rm rel}aa'}(0,\ul x;0,\ul x')
+\partial_a\partial'{}_{a'}G_{{\rm rel}00'}(0,\ul x;0,\ul x').
\eqno(5.2.11)
$$
Substituting (5.2.8) into (5.2.11), one gets
$$
K_{aa'}(\ul x;\ul x')=-{2\over \pi^2|\ul x-\ul x'|^4}
\Big[\delta_{aa'}-2{(x-x')_a(x-x')_{a'}\over |\ul x-\ul x'|^2}\Big].
\eqno(5.2.12)
$$

The Green operator (5.2.8) has the explicit Fourier representation 
$$
\eqalignno{
G_{{\rm rel}ij'}(x;x')&=
-{1\over 2}\int {d^3k\over (2\pi)^3}{1\over k}\exp(\sqrt{-1}\ul k\cdot(\ul x-\ul x'))
\Big\{\delta_{ij'}\exp(-k|x^0-x'^{0'}|)~~~~~~~~~&(5.2.13)\cr
&\phantom{=-{1\over 2}\int {d^3k\over (2\pi)^3}{1\over k}}
-\big[\delta_{ij'}-2\delta_{i0}\delta_{0'j'}\big]
\exp(-k|x^0+x'^{0'}|)\Big\}.&\cr
}
$$
Inserting (5.2.13) into (5.2.11), one gets
$$
K_{aa'}(\ul x;\ul x')=
\int {d^3k\over (2\pi)^3}\exp(\sqrt{-1}\ul k\cdot(\ul x-\ul x'))
\Big(\delta_{aa'}-{k_a k_{a'}\over k^2}\Big)k.
\eqno(5.2.14) 
$$

It is now easy to compute the action $S(a,\tau)$ in Fourier representation.
Setting 
$$
a_a(\ul k)=\int d^3x\exp(-\sqrt{-1}\ul k\cdot\ul x)a_a(\ul x)
\eqno(5.2.15)
$$
and substituting (5.2.14) into (4.2.16), one gets through a simple calculation
$$
S(a,\tau)={\sqrt{-1}\over 4\pi}
\int {d^3k\over(2\pi)^3}a_a(\ul k)a_b(-\ul k)\Big[\tau_2
\Big(\delta_{ab}-{k_a k_b\over k^2}\Big)k
+\tau_1\epsilon_{abc}k_c\Big].
\eqno(5.2.16) 
$$

Inserting (5.2.16) into (5.2.7) and plugging the result in (5.2.6), we find that 
$Z_{\hat 1}(\hat A,\tau)$, as given by (5.2.6), has precisely 
the same form as the generating functional of the correlators of the 
symmetry current of the 3 dimensional conformal field theory on $S^3$ 
corresponding to the large $N$ limit of a 3 dimensional field theory 
of $N$ fermions with $\U(1)$ symmetry, computed by Witten in \ref{1}. 
This indicates that Witten's conjecture is correct in this particular case. 
However, to be completely sure that it holds for more general 
$3$--folds $\hat M$ satisfying (5.1.3), we have to test it in other ways.

To perform the test, one needs an explicit expression of $Z_{\hat P}(\hat A,\tau)$, 
which is not readily available in general. The test is further complicated by the fact 
that expression (5.2.2) is not fully explicit for reasons explained above. 
A possible strategy to overcome these difficulties consists in making a reasonable 
hypothesis on the content of the set ${\eul M}(\hat P)$ and imposing that 
$Z_{\hat P}(\hat A,\tau)$, as given by (5.2.2), satisfies (5.2.1). In this way, 
on one hand one tests the conjecture, on the other one obtains various conditions 
which limit the validity of the hypothesis made. This is the strategy that we shall 
follow in the rest of this section. Before proceeding further, however, one should 
keep in mind that the generating functional $Z_{\hat P}(\hat A,\tau)$ 
has no natural normalization a priori. 
So, in practice, (5.2.1) holds in general only up to factors independent from 
$\hat A$ but possibly dependent on $\tau$. On the other hand, this the best one can 
hope to find out using (5.2.2), since the overall normalization of the gauge theory 
partition functions $Z^M_{\hat P}(\hat A,\tau)$ is arbitrary. 

In practice, the strategy described can be implemented only for some particularly 
simple choice of the $3$--fold $\hat M$ with conformal structure $\hat\gamma$ 
and compatible spin structure $\hat\sigma$.
Suppose that there is a distinguished $4$--fold $M$ with $\partial M=\hat M$
with the following properties valid for a broad class of data
$\hat\gamma$, $\hat\sigma$. 

\item{$i$)} 
There exists an asymptotically hyperbolic Einstein metric $g_{AHE}$ of $M$ 
with a double pole at $\hat M$ having $\hat\gamma$ as conformal infinity. 

\item{$ii$)} 
There exists a spin structure $\sigma$ of $M$ subordinated to some 
conformal compactification $g$ of $g_{AHE}$ extending $\hat\sigma$.

\item{$iii$)} 
Every bundle $\hat P\in\Princ(\hat M)$ is extendable to a bundle 
$P\in\Princ(M)$. 

\noindent
For simplicity, we assume that $M$ has no connected components without boundary.
This setting is the simplest generalization of that of the $3$--sphere 
$S^3$ just treated. (Further examples of such situation will be illustrated in 
the next section). It is then conceivable that ${\eul M}(\hat P)=\{M\}$. 
Assuming this, (5.2.2) yields the statement
$$
Z_{\hat P}(\hat A,\tau)=Z^M_{\hat P}(\hat A,\tau),
\eqno(5.2.17)
$$ 
where $Z^M_{\hat P}(\hat A,\tau)=Z^M_{P_\partial}(A_\partial,\tau)$
is given by (5.2.5). As a partial test of the conjecture, we shall check
whether and under which conditions $Z^M_{P_\partial}(A_\partial,\tau)$
fulfills (5.2.1) with $Z_{\hat P}(\hat A,\tau)$ replaced by 
$Z^M_{P_\partial}(A_\partial,\tau)$.  

\vskip .3cm {\it  5.3. Conditions on the $4$--fold $M$}

In the standard boundaryless case, when $b_2>0$, Abelian duality requires that 
the following conditions are met (cf. eq. (4.3.1), (4.3.2)) \ref{12--15}. 

\item{$i$)} The intersection matrix $Q$ is unimodular
$$
\det Q=\pm1.
\eqno(5.3.1)
$$

\item{$ii$)} The modified Hodge matrix 
$$
\tilde H=Q^{-1}H
\eqno(5.3.2)
$$
satisfies 
$$
\tilde H^2=1.
\eqno(5.3.3)
$$

\noindent
These have to be satisfied also in the present context in order 
$Z^M_{P_\partial}(A_\partial,\tau)$ to satisfy (5.2.1a), 
unless we are willing to envisage some sort of completely new mechanism.
Now, we shall show that these conditions cannot be fulfilled on an oriented compact
$4$--fold $M$ with boundary with $b_2>0$.

Let $\{F_{\rm abs}{}^r|\,r=1,\ldots,b_2\}$, $\{F_{{\rm rel}r}|\,r=1,\ldots,b_2\}$ be 
reciprocally dual bases of the spaces $\Harm^2_{\rm abs}(M)$, $\Harm^2_{\rm rel}(M)$,
respectively. Then,
$$
\int_M F_{\rm abs}{}^r\wedge F_{{\rm rel}s}=\delta^r{}_s.
\eqno(5.3.4)
$$
By the isomorphism (3.3.2), there is an invertible $b_2\times b_2$ 
matrix $H$ such that
$$
\star F_{{\rm rel}r}=H_{rs}F_{\rm abs}{}^s.
\eqno(5.3.5)
$$
Define a $b_2\times b_2$ matrix $Q$ by either relations
$$
\eqalignno{
&\int_M F_{\rm abs}{}^r\wedge F_{\rm abs}{}^s=(H^{-1}QH^{-1})^{rs},&(5.3.6a)\cr
&\int_M F_{{\rm rel}r}\wedge F_{{\rm rel}s}=Q_{rs}.&(5.3.6b)\cr
}
$$
Now, assume that the matrix $Q$ is invertible. The difference $F_{\rm abs}{}^r
-Q^{-1rs}F_{{\rm rel}s}$ is harmonic and orthogonal to all 
the $F_{\rm abs}{}^r$, by (5.3.4)--(5.3.6). 
By the Hodge decomposition theorem (3.2.3b), it follows then that 
$$
F_{\rm abs}{}^r=Q^{-1rs}(F_{{\rm rel}s}+dh_s),
\eqno(5.3.7)
$$
for certain $1$--forms $h_r\in\Omega^1(M)$. From (5.3.5), it follows that
$$
\eqalignno{
&\int_M F_{\rm abs}{}^r\wedge \star F_{\rm  abs}{}^s=H^{-1rs},&(5.3.8a)\cr
&\int_M F_{{\rm rel}r}\wedge \star F_{{\rm rel}s}=H_{rs}.&(5.3.8b)\cr
}
$$
Substituting (5.3.7) in (5.3.8a) and using (5.3.8b),
it is easy to show that
$$
H=QH^{-1}Q-R, 
\eqno(5.3.9)
$$
where $R$ is the $b_2\times b_2$ matrix
$$
R_{rs}=\int_M dh_r\wedge \star dh_s.
\eqno(5.3.10)
$$
In order $(Q^{-1}H)^2=1$, this matrix must vanish. This can happen 
if and only if $dh_r=0$ for all $r$. From (5.3.7), it
then follows that $\Harm^2_{\rm abs}(M)=\Harm^2_{\rm rel}(M)$. In general,
such identity does not hold, unless both spaces vanish. 
This contradicts the assumption that $b_2>0$.

We conclude that, in order $Z^M_{P_\partial}(A_\partial,\tau)$ to satisfy (5.2.1a), 
we must have 
$$
H^2(M,\Bbb R)=H^2(M,\partial M,\Bbb R)=0.
\eqno(5.3.11)
$$
From (5.3.11), we have therefore
$$
H^2(M,\Bbb Z)=\Tor H^2(M,\Bbb Z),\quad
H^2(M,\partial M,\Bbb Z)=\Tor H^2(M,\partial M,\Bbb Z).
\eqno(5.3.12)
$$
By the isomorphisms (2.2.2), (2.2.4), we conclude that
$$
\Princ(M)=\Princ_0(M),\quad
\Princ(M,\partial M)=\Princ_0(M,\partial M).
\eqno(5.3.13)
$$
Thus, all (relative) principal bundles on $M$ are flat. Note that this conclusion is 
consistent with (5.1.8) and the holographic formula (5.2.3).

\vskip .3cm {\it  5.4. Final simplified 
\it expression of $Z^M_{P_\partial}(A_\partial,\tau)$}

From (5.1.8), (5.3.13), we are allowed to chose the reference connections $A_c$, 
$A_{c\partial}$ so that 
$$
F_{A_c}=0.
\eqno(5.4.1)
$$
This allows for a simplification of eq. (5.2.5).
The expression of the functional $Z^M_{P_\partial}(A_\partial,\tau)$ 
becomes in this way
$$
Z^M_{P_\partial}(A_{c\partial}+a,\tau)=t_{\rm rel}^2\,\tau_2^{b_{{\rm rel}1}/2}\,
\exp\left(\sqrt{-1}S(a,\tau)\right)\equiv {\cal Z}(a,\tau).
\eqno(5.4.2)
$$
The factor $t_{\rm rel}^2\,\tau_2^{b_{{\rm rel}1}/2}$ is irrelevant
for reasons explained at end of subsect. 5.2. So, it can be dropped.
By (5.4.2), $Z^M_{P_\partial}(A_{c\partial}+a,\tau)$ is independent from 
the bundle $P_\partial\in\Princ(\partial M)$ and the reference connection 
$A_{c\partial}\in\Conn(P_\partial)$. (5.4.2) is essentially (1.9)
Then, as is easy to see, the $\SL(2,\Bbb Z)$ 
operators $\widehat S$, $\widehat T$, $\widehat C$ defined in (5.1.1),
when acting on $Z^M_{P_\partial}(A_\partial,\tau)$, take the simpler form (1.5), 
with ${\cal Z}_{\scri T}(a)$ replaced by ${\cal Z}(a,\tau)$, and the relations 
(5.2.1) reduce into those (1.7). Note that the sum in (5.1.1a) is actually finite, 
since, by (5.1.8), $\Princ(\hat M)$ is a finite group. 

As a check, let us now verify that the gauge invariance requirement (5.1.2) is 
fulfilled by $Z^M_{P_\partial}(A_\partial,\tau)$. In terms of the functional  
${\cal Z}(a,\tau)$ of eq. (5.4.1), this takes precisely the form (1.2), 
with ${\cal Z}_{\scri T}(a)$, $d$ replaced by ${\cal Z}(a,\tau)$,
$d_\partial$. Suppose we shift the background $1$--form
$a\in\Omega^1(\partial M)$ by an exact $1$--form $d_\partial f$, where  
$f\in\Omega^0(\partial M)$. To see what happens, let us go back to
the boundary value problem (4.2.1). Under the shift, the $1$--form
$\omega\in\Omega^1(M)$ gets replaced by $\omega+\varpi+d\varphi$, where 
the relative harmonic 1--form $\varpi\in\Harm_{\rm rel}^1(M)$ is arbitrary
and the $0$--form $\varphi\in\Omega^0(M)$ satisfies
$d^*d\varphi=0$, with $\iota_\partial{}^*\varphi=f$. 
Indeed, by an argument similar to that leading to (4.2.3), one shows 
that the solution of this boundary value problem, 
if any, is unique if $H^0(M,\partial M,\Bbb R)=0$. 
This is indeed the case by a general theorem, since $M$ has no connected components 
without boundary. 
The vanishing of $b_{{\rm rel}0}$ ensures the existence and 
uniqueness of a relative Green operator 
$L_{\rm rel}$ for the Hodge Laplacian $\Delta$ on $\Omega_{\rm rel}^0(M)$. 
Using $L_{\rm rel}$, one can compute $\varphi$ in terms 
of $f$ through the general identity (3.1.5):
$\varphi(x)=-\oint_{\partial M}f\wedge
\star\iota_\partial{}^*j(n)dL_{{\rm rel}x}$.
It follows that under the shift of $a$ by $d_\partial f$, $d\omega$ is invariant.
In our calculation, $\omega$ is the connection ${\cal A}$ (cf. subsect. 4.1). 
Hence, under the shift, the gauge curvature
$d{\cal A}$ is invariant. Since the action $S(a,\tau)=S({\cal A},\tau)$
depends on $a$ through $d{\cal A}$, we have 
$$
S(a+d_\partial f,\tau)=S(a,\tau).
\eqno(5.4.3)
$$
By (5.4.1), ${\cal Z}(a,\tau)$ satisfies (1.2) as required.

\vskip .3cm {\bf 6. Discussion and examples}
\vskip .3cm 

In the previous section, we tested Witten's holographic conjecture for the oriented 
compact $3$--folds $\hat M$ satisfying (5.1.3) for which there exists a 
distinguished oriented compact $4$--fold $M$ with $\partial M=\hat M$ 
with the following properties. 

\item{$i$)} Every bundle $\hat P\in\Princ(\hat M)$ is extendable to $M$. 

\item{$ii$)} For every conformal structure $\hat\gamma$ and 
compatible spin structure $\hat\sigma$ of $\hat M$ varying 
in some broad class, there exists an asymptotically hyperbolic Einstein metric 
$g_{AHE}$ of $M$ with a double pole at $\hat M$ having $\hat\gamma$ 
as conformal infinity and there exists a spin structure $\sigma$ of $M$ 
subordinated to some conformal compactification $g$ of $g_{AHE}$ extending  
$\hat\sigma$. 

\noindent
We found that the validity of the conjecture requires that $M$ satisfies (5.3.11). 
Next, we shall look for sufficient conditions for such an $M$ to exist. 
Before proceeding further, however, it is necessary to remark that the conditions,
which we shall find, are {\it not} necessary for the validity of the 
conjecture, which may hold true for a more general topological setting than the one 
considered here, and are ultimately motivated only by simplicity.

We tackle the problem by the following strategy. We consider an appropriate set 
of pairs $(\hat M,M)$ with $\partial M=\hat M$ and look for sufficient conditions 
for property $i$ and $ii$ to hold.

\vskip .3cm {\it  6.1. Simple admissible pairs}

A pair $(\hat M,M)$ formed by an oriented compact $3$--folds $\hat M$ 
and an oriented compact $4$--fold $M$ with $\partial M=\hat M$ is said
{\it admissible} if $\hat M$ satisfies (5.1.3), $M$ satisfies (5.3.11)
and $M$ has no connected components without boundary.

Let $(\hat M,M)$ be an admissible pair. From (5.1.3), (5.3.11) and from Poincar\`e 
duality, we have 
$$
\eqalignno{
\vphantom{\int}
&H^p(M,\partial M,\Bbb R)\cong H^{4-p}(M,\Bbb R)=0, \quad p=0,2, &(6.1.1a)\cr
\vphantom{\int}
&H^1(\partial M,\Bbb R)\cong H^2(\partial M,\Bbb R)=0. &(6.1.1b)\cr
}
$$
The vanishing of the relative cohomology space for $p=0$ is a general theorem.

As we saw above, things simplify considerably when (4.2.3) holds, i. e when 
(6.1.1a) holds also for $p=1$. When this does happen, the admissible pair
$(\hat M,M)$ is said {\it simple}.

Let $(\hat M,M)$ be a simple admissible pair. From (6.1.1) and the 
absolute/relative cohomology long exact sequence (2.1.5) with ${\scri S}=\Bbb R$, 
we conclude that
$$
H^p(M,\partial M,\Bbb R)\cong H^{4-p}(M,\Bbb R)=0, \quad p=0,1,2,3.
\eqno(6.1.2)
$$
Thus, $M$ has trivial real (relative) cohomology and is so $M$ is acyclic.
From (6.1.1b), we have further that 
$$
H^{3-p}(\partial M,\Bbb R)\cong H^p(\partial M,\Bbb R)\cong H^p(M,\Bbb R)
\quad p=0,~1,
\eqno(6.1.3)
$$
with the three terms vanishing for $p=1$. This relation implies in particular that
$M$ is connected if and only if $\partial M$ is. 

Below, we shall concentrate on simple admissible pairs, for simplicity.

\vskip .3cm {\it  6.2. Sufficient condition for property $i$ to hold}

Let $(\hat M,M)$ be a simple admissible pair. There is an obvious sufficient condition 
for the validity of property $i$. Recall that, by the isomorphism (2.2.2), 
$\Princ(\hat M)\cong H^2(\hat M,\Bbb Z)$. If 
$$
H^2(\hat M,\Bbb Z)=0,
\eqno(6.2.1) 
$$
then 
$$
\Princ(\hat M)=0,
\eqno(6.2.2) 
$$
i. e. $\hat M$ supports only the trivial bundle $\hat 1$. $\hat 1$ is obviously 
extendable to the $4$--fold $M$ (cf. subsect. 2.5). So property $i$ holds trivially. 
Note that,  by Poincar\'e  duality, (6.2.1) implies and, thus, is compatible with 
(5.1.3). 

If $\hat M$ is connected, it follows from (6.2.1) that $\hat M$ is an integer homology 
$3$--sphere. In that case, $M$ is a rational homology $4$--ball, for reasons
explained at the end of subsect. 6.1.

The pairs $(\hat M, M)$ where $\hat M$ is an integer homology 
$3$-sphere and $M$ is a contractible $4$--fold constitute an interesting 
class of simple admissible pairs of the type just described 
having property $i$, since, cohomologically, they 
are indistinguishable from the pair $(S^3,B^4)$, for which Witten's conjecture was 
explicitly checked in subsect. 5.2. The integer homology $3$--spheres $\hat M$ 
bounding a contractible $4$--fold form a class of $3$--folds that has been 
extensively studied. Such a class includes a subset of the well--known Brieskorn 
$3$--folds, which now we describe. Let $p,~q,~r\in\Bbb N$ be relative prime.
The Brieskorn homology $3$--sphere $\Sigma(p,q,r)$ is by definition
$$
\Sigma(p,q,r)=\{z\,|\,z\in\Bbb C^3,~|z|^2=1,~(z^1)^p+(z^2)^q+(z^3)^r=0\}.
\eqno(6.2.3)
$$
$\Sigma(p,q,r)$ is a naturally oriented smooth compact $3$--fold. If the 
triple $(p,q,r)$ belongs to the following list
$$
\eqalignno{
\vphantom{\int}
&(p,ps-1,ps+1),\quad \hbox{$p$ even, $s$ odd}&(6.2.4a)\cr
\vphantom{\int}
&(p,ps\pm 1,ps\pm 2),\quad \hbox{$p$ odd}&(6.2.4b)\cr
\vphantom{\int}
&(2,2s\pm 1,4(2s\pm 1)+2s\mp 1),\quad \hbox{$s$ odd}&(6.2.4c)\cr
\vphantom{\int}
&(3,3s\pm 1,6(3s\pm 1)+3s\mp 2),&(6.2.4d)\cr
\vphantom{\int}
&(3,3s\pm 2,6(3s\pm 2)+3s\mp 1),&(6.2.4e)\cr
}
$$
with $p,~s\in\Bbb N$, then $\Sigma(p,q,r)$ bounds a a contractible $4$--fold 
$B(p,q,r)$ \ref{23,24}. So, the pairs $(\Sigma(p,q,r),B(p,q,r))$ with 
$(p,q,r)$ belonging to the list (6.2.4) are simple admissible and 
have property $i$.

\vskip .3cm {\it  6.3. Sufficient condition for property $ii$ to hold}

To the best of our knowledge, there are no general theorems ensuring that a simple 
admissible pair $(\hat M, M)$ admits an asymptotically hyperbolic Einstein metric 
$g_{AHE}$ of $M$ with a double pole at $\hat M$ having a given conformal structure 
$\hat\gamma$ of $\hat M$ as conformal infinity. The matter has not been settled 
completely yet even for the pair $(S^3,B^4)$. The analysis of specific examples 
indicates that the general solubility of the existence problem for given $\hat M$ 
depends on both the set ${\eul C}$ of conformal structures of $\hat M$ considered and 
the topology of $M$. Partial results can be found in \ref{25}, which we summarize 
next.

Let $M$ be an oriented compact $4$--fold with boundary. 
Let ${\eul C}(\partial M)$ be the set of the non negative conformal structures 
$\gamma_\partial$ of $\partial M$, i. e. containing a non flat representative metric 
$g_\partial$ with non negative scalar curvature. Let $\bar{\eul E}_{AHE}(M)$ be the 
set of the asymptotically hyperbolic Einstein metrics $g_{AHE}$ of $M$ with a 
double pole at $\partial M$ whose conformal infinity is contained in  
${\eul C}(\partial M)$ modulo the action of the diffeomorphisms of $M$ inducing 
the identity on $\partial M$. ${\eul C}(\partial M)$, $\bar{\eul E}_{AHE}(M)$ 
have natural structures of smooth Banach manifolds.
Let $\Pi_M:\bar{\eul E}_{AHE}(M)\rightarrow {\eul C}(\partial M)$ 
the natural map that associates to any $g_{AHE}\in\bar{\eul E}_{AHE}(M)$ its conformal 
infinity $\Pi_M(g_{AHE})\in{\eul C}(\partial M)$. In \ref{25}, it is shown that,
if the sequence
$$
H^1(\partial M,\Bbb R)\longrightarrow H^2(M,\partial M,\Bbb R)\longrightarrow 0 
\eqno(6.3.1)
$$
induced by the inclusion map $\iota_\partial$ is exact, then 
$\Pi_M$ is a proper Fredholm map of index $0$. It is then 
possible to define an integer valued degree of the map $\Pi_M$:
$$
\deg\Pi_M=\sum_{g\in\Pi_M{}^{-1}(\gamma_\partial)}(-1)^{\ind g}\in\Bbb Z.
\eqno(6.3.2)
$$ 
Here, $\gamma_\partial\in{\eul C}(\partial M)$ is a structure, whose choice 
is immaterial. $\ind g$ is the index of the elliptic operator $L_g$ obtained by 
linearization of the Einstein equations at $g$ and is defined as the maximal dimension 
of the subspaces of the domain of $L_g$ on which $L_g$ is a negative definite
bilinear form with respect to the standard $L^2$ inner product. Then, if 
$$
\deg\Pi_M\not=0,
\eqno(6.3.3)
$$
$\Pi_M$ is surjective. In that case, for every 
$\gamma_\partial\in{\eul C}(\partial M)$,
there is a metric $g_{AHE}\in\bar{\eul E}_{AHE}(M)$ having $\gamma_\partial$ as
conformal infinity.

From the above considerations, it seems natural to set 
$$
{\eul C}={\eul C}(\hat M),
\eqno(6.3.4) 
$$
for any simple admissible pair $(\hat M, M)$. It is remarkable that, since $M$ 
satisfies (5.3.11), (6.3.1) is exact and, so, $\deg\Pi_M$ can be defined. So, for 
a simple admissible pair $(\hat M, M)$, there is an asymptotically hyperbolic Einstein 
metric $g_{AHE}$ for any given conformal structure $\hat\gamma\in{\eul C}$, 
if (6.3.3) holds. 
In \ref{26}, it was shown that the solubility of the above existence problem implies 
that $H^1(M,\partial M,\Bbb Z)=0$. This condition implies (4.2.3), which holds 
anyway, since $(\hat M, M)$ is simple by assumption. 

Recall that a Riemannian oriented manifold $X$ can support spin structures if and 
only if its 2nd Stiefel--Whitney class $w_2(X)$ vanishes and that, in that case, 
its spin structures are parametrized by $H^1(X,\Bbb Z_2)$ \ref{27}.
Recall also that, for every oriented $3$--fold $X$, $w_2(X)=0$ \ref{27}.
For every $3$--fold $\hat M$ satisfying (5.1.3) and every conformal structure 
$\hat\gamma\in{\eul C}(\hat M)$, $\hat M$ supports precisely one spin structure 
$\hat\sigma$ compatible with $\hat\gamma$, since $w_2(\hat M)=0$ and
$H^1(\hat M,\Bbb Z_2)=0$, by (5.1.4) and the universal coefficient theorem. 
Let $(\hat M, M)$ be a simple admissible pair such that (6.3.3) holds and that 
$$
w_2(M)=0.
\eqno(6.3.5) 
$$ 
Let $\hat\gamma\in{\eul C}(\hat M)$ and let $g_{AHE}\in\bar{\eul E}_{AHE}(M)$ be an 
asymptotically hyperbolic Einstein metric of $M$ having $\hat\gamma$ as conformal 
infinity. Let $\hat\sigma$ be the unique spin structure of $\hat M$ compatible 
with $\hat\gamma$. Let $g$ be a conformal compactification of $g_{AHE}$ and 
$\sigma$ be a spin structure of $M$ subordinated to $g$. Then, $\sigma$ extends 
automatically $\hat\sigma$, since $\sigma$ induces a spin structure on $\hat M$
which necessarily coincides with $\hat\sigma$. 
We conclude that an admissible pair $(\hat M, M)$ has property $ii$, if it satisfies 
(6.3.3) and (6.3.5).

The simple admissible pairs $(\Sigma(p,q,r),B(p,q,r))$ with $(p,q,r)$ in the 
list (6.2.4) satisfy (6.3.5), since $B(p,q,r)$ is contractible. It would be 
interesting to determine which of these pairs satisfy also (6.3.3). Unfortunately, 
this is not known presently to the best of our knowledge. 
 
\vskip.6cm
\par\noindent
{\bf Acknowledgments.} I am grateful to D. Anselmi, F. Bastianelli, M. Matone, 
R. Stora and especially E. Witten for helpful discussions. I thank P. Lisca, 
R. Kirby, R. Graham and M. Anderson for correspondence.

\vskip.6cm
\centerline{\bf REFERENCES}
\vskip.6cm

\item{[1]}
E.~Witten,
``SL(2,Z) action on three-dimensional conformal field theories 
with Abelian symmetry'', 
arXiv:hep-th/0307041.

\item{[2]}
A.~Kapustin and M.~J.~Strassler,
``On mirror symmetry in three dimensional Abelian gauge theories'',
JHEP {\bf 9904} (1999) 021,
arXiv:hep-th/9902033.

\item{[3]}
M.~Henningson,
``Extended superspace, higher derivatives and SL(2,Z) duality'',
Nucl.\ Phys.\ B {\bf 458} (1996) 445, 
arXiv:hep-th/9507135.

\item{[4]}
T.~Appelquist and R.~D.~Pisarski,
``Hot Yang-Mills theories and three-dimensional QCD'',
Phys.\ Rev.\ D {\bf 23} (1981) 2305.

\item{[5]}
R.~Jackiw and S.~Templeton,
``How superrenormalizable interactions cure their infrared divergences'',
Phys.\ Rev.\ D {\bf 23} (1981) 2291.

\item{[6]}
S.~Templeton,
``Summation of dominant coupling constant logarithms in QED in three dimensions'',
Phys.\ Lett.\ B {\bf 103} (1981) 134.

\item{[7]}
T.~Appelquist and U.~W.~Heinz,
``Three dimensional O(N) theories at large distances'',
Phys.\ Rev.\ D {\bf 24} (1981) 2169.

\item{[8]}
V.~Borokhov, A.~Kapustin and X.~K.~Wu,
``Topological disorder operators in three-dimensional conformal field theory'',
JHEP {\bf 0211} (2002) 049,
arXiv:hep-th/0206054.

\item{[9]}
V.~Borokhov, A.~Kapustin and X.~K.~Wu,
``Monopole operators and mirror symmetry in three dimensions'',
JHEP {\bf 0212} (2002) 044,
arXiv:hep-th/0207074.

\item{[10]}
R.~G.~Leigh and A.~C.~Petkou,
``SL(2,Z) action on three-dimensional CFTs and holography'',
arXiv:hep-th/0309177.

\item{[11]}
C.~Vafa and E.~Witten,
``A Strong coupling test of S duality'',
Nucl.\ Phys.\ B {\bf 431} (1994) 3, 
arXiv:hep-th/9408074.

\item{[12]}
E.~Witten,
``On S duality in Abelian gauge theory'',
Selecta Math.\  {\bf 1} (1995) 383, 
arXiv:hep-th/9505186.

\item{[13]}
M.~Alvarez and D.~I.~Olive,
``The Dirac quantization condition for fluxes on four-manifolds'',
Commun.\ Math.\ Phys.\  {\bf 210} (2000) 13,
arXiv:hep-th/9906093.

\item{[14]}
D.~I.~Olive and M.~Alvarez,
``Spin and Abelian electromagnetic duality on four-mani\-folds'',
Commun.\ Math.\ Phys.\  {\bf 217} (2001) 331,
arXiv:hep-th/0003155.

\item{[15]}
R.~Zucchini,
``Abelian duality and Abelian Wilson loops'', 
Commun.\ Math.\ Phys.\  {\bf 242} (2003) 473,
arXiv:hep-th/0210244.

\item{[16]}
R.~Bott and L.~Tu,
``Differential forms in algebraic topology'',
Graduate Texts in Mathematics 82, Springer Verlag, New York (1982).

\item{[17]}
G.~E.~Bredon,
``Sheaf Theory'',
Graduate Texts in Mathematics 170, Springer Verlag, New York (1997).

\item{[18]}
P.~B.~Gilkey, J.~V.~Leahy and J.~Park
``Spinors, spectral geometry and Riemannian submersions'',
Lecture Notes Series {\bf 40}, Research Institute of Mathematics, Global Analysis 
Research Center, Seoul National University (1998), 
available at the EMIS server http://www.emis.de/monographs/GLP/index.html.

\item{[19]}
P.~B.~Gilkey,
``Invariance theory, the heat equation and the Atiyah-Singer index theorem'',
Publish or Perish, Wilmington (1984).

\item{[20]}
E.~Witten,
``Anti-de Sitter space and holography'',
Adv.\ Theor.\ Math.\ Phys.\  {\bf 2} (1998) 253,
arXiv:hep-th/9802150.

\item{[21]}
R.~Dijkgraaf, J.~M.~Maldacena, G.~W.~Moore and E.~Verlinde,
``A black hole farey tail'',
arXiv:hep-th/0005003.

\item{[22]}
C.~Fefferman and C.~R.~Graham,
``Conformal invariants'', in ``The mathematical heritage of Elie Cartan'',
Asterisque (1985) 95.

\item{[23]}
R.~J.~Stern,
``Some more Brieskorn spheres which bound contractible manifolds'',
Notices \ Amer.\ Math.\ Soc. 25 {\bf A448} (1978), abstract 78T--G75, problem 4.2.

\item{[24]}
A.~Casson and J.~Harer, 
``Some homology lens spaces which bound rational homology balls'',  
Pacific \ J.\ Math. {\bf 96} (1981), no. 1, 23.

\item{[25]}
M.~T.~Anderson,
``Einstein metrics with prescribed conformal infinity on 4-manifolds'',
arXiv:math.DG/0105243.

\item{[26]}
E.~Witten and S.~T.~Yau,
``Connectedness of the boundary in the AdS/CFT correspondence'',
Adv.\ Theor.\ Math.\ Phys.\  {\bf 3} (1999) 1635,
arXiv:hep-th/9910245.

\item{[27]}
H,~Blaine Lawson and M.~-L.~Michelsohn,
``Spin geometry'', Princeton University Press, Princeton (1989).
\bye